\documentclass[11pt]{article}
\usepackage{times}
\usepackage{textgreek}
\usepackage{geometry}
\usepackage{eurosym}
\usepackage[utf8]{inputenc}
\usepackage{enumitem,amssymb}
\usepackage{ragged2e}
\newlist{thematic}{itemize}{8}
\setlist[thematic]{label=$\square$}
\usepackage{pifont}
\usepackage[utf8]{inputenc}

\usepackage[english]{babel}
\usepackage{amsfonts}
\usepackage{wrapfig}
\usepackage{hyperref}
\usepackage{graphbox}
\usepackage{multirow}
\usepackage{makecell}
\usepackage{amsmath,amssymb,gensymb}
\usepackage{mathtools}
\usepackage[sort&compress,numbers]{natbib}
\usepackage{url}
\usepackage{siunitx}
\usepackage{xspace}
\usepackage{microtype}
\usepackage{subfig}
\usepackage[useregional]{datetime2}
\usepackage{graphicx}
\usepackage{lineno}
\usepackage{siunitx}
\usepackage{xspace}

\newcommand{\NEW}{NEXT-White}

\newcommand{\tz}{\ensuremath{t_0}}
\newcommand{\stwo}{\ensuremath{S_2}}
\newcommand{\sone}{\ensuremath{S_1}}

\newcommand{\mbb}{\ensuremath{m_{\beta\beta}}}

\newcommand{\ckky}{\ensuremath{\rm counts~keV^{-1}~kg^{-1}~yr^{-1}}}

\newcommand{\Qbb}{\ensuremath{Q_{\beta\beta}}}

\newcommand{\Tonu}{\ensuremath{T_{1/2}^{0\nu}}}

\newcommand{\RAD}{\ensuremath{^{222}}Rn}

\newcommand{\XE}{\ensuremath{{}^{136}\rm Xe}}

\newcommand{\Bapp}{Ba\ensuremath{^{++}}}

\DeclareSIUnit\c{\mbox{$c$}}
\DeclareSIUnit\magn{\mbox{$\times$}}
\DeclareSIUnit\min{min}
\DeclareSIUnit\week{week}
\DeclareSIUnit\year{yr}
\DeclareSIUnit\years{years}
\DeclareSIUnit\yr{yr}
\DeclareSIUnit\standard{std}
\DeclareSIUnit\str{sr}
\DeclareSIUnit\ppm{ppm}
\DeclareSIUnit\ppb{ppb}
\DeclareSIUnit\ppt{ppt}
\DeclareSIUnit\pe{PE}
\DeclareSIUnit\spe{SPE}
\DeclareSIUnit\ev{events}
\DeclareSIUnit\ct{counts}
\DeclareSIUnit\neutron{\mbox{$n$}}
\DeclareSIUnit\smp{samples}
\DeclareSIUnit\Sample{S}
\DeclareSIUnit\ch{ch}
\DeclareSIUnit\hit{hit}
\DeclareSIUnit\hits{hits}
\DeclareSIUnit\bin{(\mbox{5-PE}~bin)}
\DeclareSIUnit\sgm{\mbox{$\sigma$}}
\DeclareSIUnit\rms{RMS}
\DeclareSIUnit\keVr{\mbox{keV$_{\rm nr}$}}
\DeclareSIUnit\keVee{\mbox{keV$_{e{\rm e}}$}}
\DeclareSIUnit\ph{photon}
\DeclareSIUnit\pes{pes}
\DeclareSIUnit\el{electrons}
\DeclareSIUnit\pm{PMT}
\DeclareSIUnit\inch{"}
\DeclareSIUnit\bit{bit}
\DeclareSIUnit\sample{samples}
\DeclareSIUnit\barn{barn}
\DeclareSIUnit\bara{bar}
\DeclareSIUnit\barg{barg}
\DeclareSIUnit\mlardepth{\mbox(meter~of~\LAr~depth)}
\DeclareSIUnit\Curie{Ci}
\DeclareSIUnit\psi{psi}
\DeclareSIUnit\parsec{pc}
\DeclareSIUnit\liveday{\mbox{live-days}}
\DeclareSIUnit\days{\mbox{days}}
\DeclareSIUnit\day{\mbox{day}}
\DeclareSIUnit\miles{\mbox{miles}}
\DeclareSIUnit\degreeC{\mbox{$^{\circ}$C}}
\DeclareSIUnit\electron{\mbox{$e^-$}}
\DeclareSIUnit\Euro{\mbox{\euro}}
\DeclareSIUnit\cph{cph}
\DeclareSIUnit\neq{neq}
\DeclareSIUnit\unit{unit}
\DeclareSIUnit\byte{Byte}
\DeclareSIUnit\Bq{\becquerel}

\newcommand{\Next}{\mbox{NEXT-100}}

\newcommand{\HPXeEL}{HPXe-EL}

\newcommand{\XeEnrichment}{\SI{90}{\percent}}

\newcommand{\NextTpcDiameter}{\SI{1050}{\mm}}
\newcommand{\NextTpcLength}{\SI{1300}{\mm}}
\newcommand{\NextFiducialVolume}{\SI{1.27}{\cubic\meter}}

\newcommand{\NextFiducialMass}{\SI{97}{\kg}}

\newcommand{\NextPressure}{\SI{15}{\bar}}

\newcommand{\NextNumberOfSiPM}{\num{5600}}
\newcommand{\NextNumberOfPMT}{\num{60}}

\newcommand{\ctsper}    {cts/(keV$\cdot$kg$\cdot$yr)}

\newcommand{\kgyr}      {kg$\cdot$yr}
\newcommand{\tyr}       {t$\cdot$yr}

\newcommand{\LEG}       {\textsc{Legend}}
\newcommand{\Ltwo}      {{\LEG-200}}
\newcommand{\Lthou}     {{\LEG-1000}}

\newcommand{\MJD}       {\textsc{Majorana Demonstrator}}
\newcommand{\Gerda}     {\textsc{Gerda}}
\newcommand{\GERDA}     {\textsc{GERDA}}

\newcommand{\Lngs}      {\textsc{Lngs}}

\newcommand{\be}        {\begin{equation}}
\newcommand{\ee}        {\end{equation}}

\usepackage[table]{xcolor}
\definecolor{swotS}{RGB}{226,237,143}
\definecolor{swotW}{RGB}{247,193,139}
\definecolor{swotO}{RGB}{173,208,187}
\definecolor{swotT}{RGB}{192,165,184}
\usepackage[raster]{tcolorbox}

\usepackage{geometry}
\usepackage{tikz}
\usetikzlibrary{calc}

\def\GanttHeader#1#2#3#4{%
 \pgfmathparse{(#1-#2-#3)/#4}
 \tikzset{y=7mm, task number/.style={left},
     task description/.style={text width=#3,  right, draw=none,
           xshift=#2,
           minimum height=2em},
     gantt bar/.style={draw=black, fill=swotO!60},
     help lines/.style={draw=black!30, dashed},
     x=\pgfmathresult pt
     }
  \def\totalmonths{#4}
  \node (Header) [task description] at (0,0) {\textbf{\large Tasks}};
  \begin{scope}[shift=($(Header.south east)$)]
    \foreach \x in {1,...,#4}
      \node[above] at (\x,0) {\footnotesize\x};
 \end{scope}
}
\def\Task#1#2#3#4{%
\node[task number] at ($(Header.west) + (0, -#1)$) {#1};
\node[task description] at (0,-#1) {#2};
\begin{scope}[shift=($(Header.south east)$)]
  \draw (0,-#1) rectangle +(\totalmonths, 1);
  \foreach \x in {1,...,\totalmonths}
    \draw[help lines] (\x,-#1) -- +(0,1);
  \filldraw[gantt bar] ($(#3, -#1+0.2)$) rectangle +(#4,0.6);
\end{scope}
}

\newcommand{\hl}     {\ensuremath{T_{1/2}}}
\newcommand{\nubb}   {DBD\ensuremath{0\nu}}

\newcommand{\senexp} {\ensuremath{\mathcal{E}}}
\newcommand{\senbkg} {\ensuremath{\mathcal{B}}}
\newcommand{\senroi} {\ensuremath{\text{ROI}}}

\title{\bf \huge Double Beta Decay APPEC Committee\protect\\ Report\protect\\ {\small Version 3}}

\begin{document}
\maketitle

{\large Committee members: Andrea Giuliani, J.J. Gomez Cadenas, Silvia Pascoli (Chair), Ezio Previtali, Ruben Saakyan, Karoline Sch\"affner and Stefan Sch\"onert}

\vspace{2truecm}

\begin{figure}[h]
    \centering
    \includegraphics[width=0.8\textwidth]{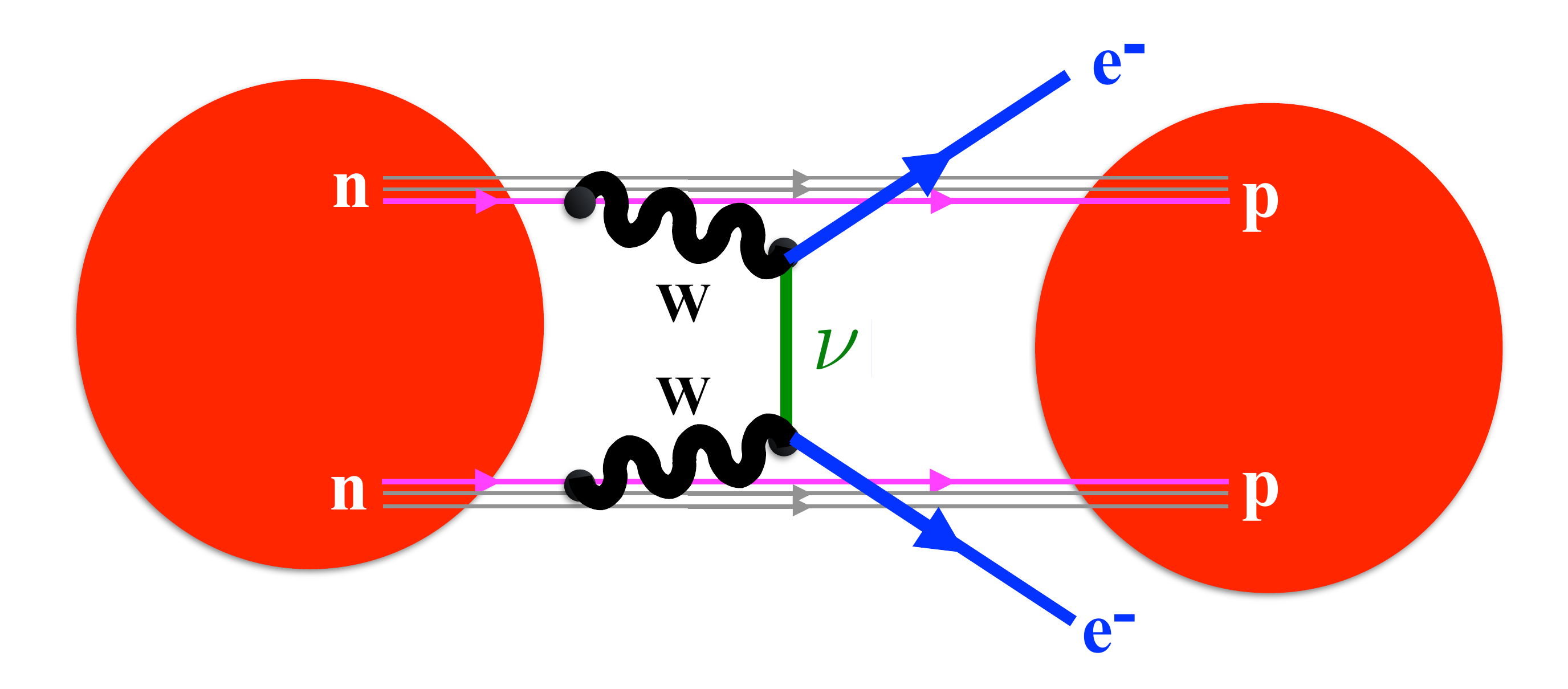}
    \caption{Schematic view of neutrinoless double beta decay.}
    \label{fig:betabetaimage}
\end{figure}

\newpage

\centerline{\bf \Large Executive Summary}

\vspace{4truemm}
This report presents a roadmap document prepared by the Double Beta Decay APPEC Committee for the APPEC Scientific Advisory Committee (SAC). The former committee received a mandate in December 2018 to provide advice to APPEC on the future neutrinoless double beta decay experimental program in Europe. After feedback from SAC and input from the community, this 
%
report gives an overview of the existing, planned and proposed technologies for neutrinoless double beta decay, their discovery potential and technical challenges making a critical examination of resources and schedules. Theoretical questions concerning the particle and nuclear physics aspects of neutrinoless double beta decay are also discussed. 

 The report provides a list of recommendations that are aimed to ensure Europe's leading role in this area of physics.


\vspace{0.1truecm}

{\em Recommendation 1. The search for neutrinoless double beta decay
is a top priority in particle and astroparticle physics, as this process provides the most sensitive test of lepton number violation. }

The discovery of neutrino masses and mixing, implied by neutrino oscillations, is so far the only particle physics evidence of physics beyond the Standard Model (SM). It has opened new key questions, among which establishing the nature of neutrinos is arguably the most important. The latter is intrinsically related to the conservation of lepton number, which is related to the fundamental symmetries of nature, the origin of neutrino masses in theories beyond the Standard Model and the generation of the observed matter-antimatter asymmetry in the Universe via the leptogenesis mechanism. Generically, neutrinoless double beta decay (DBD0$\nu$) is the most sensitive probe of lepton number violation.
The 3 light massive neutrinos, if they are of Majorana type, induce this process with half-lives which may be at reach in current and future experiments. The theoretical predictions depend on the values of neutrino masses, whether they are with normal ($m_1<m_2<m_3$) or inverted ($m_3<m_1<m_2$) ordering, and on the CP violating phases.
 Other lepton number violating processes, e.g. sterile neutrinos, in extensions of the SM give also a contribution to neutrinoless double beta decay at a level which, in specific models, may be relevant for current and future experiments.

\vspace{2truemm}

{\em Recommendation 2. A sustained and enhanced support of the European experimental programme is required to maintain the leadership in the field,  exploiting the broad range of expertise and infrastructure and fostering existing and future international collaborations.}

Key technologies in the search for neutrinoless double beta decay have been conceived, developed and demonstrated in Europe: germanium diodes operated in liquid argon with high energy resolution and multi-site event rejection; pure and scintillating bolometers capable of studying at least four different isotopes with high energy resolution and particle identification; gaseous Xenon TPC capable of combining good energy resolution with topological reconstruction of the events and, in future, final state identification. Other developments include the use of room temperature semiconductor detectors and exploiting synergies between dark matter and double beta decay searches in Xe-based experiments. The most promising approaches based on these technologies should be supported  to ensure European leadership and at the same time to foster international cooperation given the highly international nature of the field. 
The potential for European laboratories to host next-generation double beta decay experiments should be exploited.

\vspace{2truemm}

{\em Recommendation 3. A multi-isotope program exploiting different technologies at the highest level of sensitivity should be supported in Europe in order to mitigate the risks and to extend the physics reach of a possible discovery.}

The current objective of the experimental search for neutrinoless double beta decay  is to fully explore the inverted ordering region of the neutrino mass pattern. Several proposed next-generation projects aim at this goal. Some of them can in principle fully cover this region and detect DBD0$\nu$ even in case of normal ordering, provided that the lightest neutrino mass is larger than 10--20~meV. 

Future projects can be broadly classified into two categories: experiments using a fluid-embedded DBD0$\nu$ source (featured by large sensitive masses and relatively easy scalability) and experiments using a crystal-embedded DBD0$\nu$ source (featured by high energy resolution and efficiency). In the first class we have Xe-based TPC projects like nEXO (evolution of the completed EXO-200), NEXT-HD (evolution of the imminent NEXT-100), PandaX-III-1t (evolution of the foreseen PandaX-III-200) and the DARWIN dark matter detector.  This class includes also experiments which dissolve the source in a large liquid-scintillator matrix exploiting existing infrastructures such as  KamLAND2-Zen (evolution of the current KamLAND-Zen-800) and SNO+-phase-II (evolution of the imminent SNO+-phase-I). In the second class we have experiments based on germanium diodes such as LEGEND-1000 (evolution of the current GERDA and MAJORANA and of the planned LEGEND-200) and those which exploit the bolometric technique, such as the multi-step AMoRE program (AMoRE-I and AMoRE-II, which represent the evolution of the current AMoRE pilot), and CUPID, which is based on a large  experience acquired by CUORE and the CUPID-Mo and CUPID-0 demonstrators,  which are all collecting data. 
In addition, the SuperNEMO tracker-calorimeter approach remains the best way to explore a signal above 50 meV with multiple isotopes combined with a full topological reconstruction of the final state events, and further R\&D may be able to push this further into the inverted ordering region.

In this rich landscape, the most prominent projects with a strong European component are CUPID, \Lthou~ and NEXT-HD. 
These projects feature a planned 3$\sigma$ discovery sensitivity that, for some matrix element calculations, reaches below 20 meV for $m_{\beta \beta}$, and will therefore ensure that Europe remains in a forefront position in this highly competitive international endeavour. The projects will study three different isotopes ($^{100}$Mo, $^{76}$Ge and $^{136}$Xe respectively) using different technological approaches, offering a strong complementarity which is critical in such a challenging area of research.

A multi-technology approach is necessary to mitigate risks of individual experiments and to corroborate their findings given the experimental challenges posed. The use of multiple isotopes may allow the underlying mechanism behind the process to be identified sheding light on whether it is mediated by light neutrino masses or other more exotic physics.

\vspace{2truemm}

 {\em Recommendation 4. A program of R\&D should be devised on the path towards the meV scale for the effective Majorana mass parameter.}

If the neutrino mass ordering is normal and the lightest neutrino mass is below 10-20 meV, only experiments with near-zero background in the tens of tons scale will have a chance to detect neutrinoless double beta decay assuming the light neutrino mass mechanism is dominant. This poses a formidable challenge for existing technologies. However, extensions of  present approaches or totally new ideas could in principle achieve this  target if underpinned by an adequate R\&D program. These R\&D activities should be supported in order to sustain the longer term future of double beta decay research. The required large scale enrichment remains by itself a major challenge, which could be addressed  by developing a dedicated international facility as part of the research program.

\vspace{2truemm}

{\em Recommendation 5. The European underground laboratories should provide the required space and infrastructure for next generation double beta decay experiments. A strong level of coordination is required among European laboratories for radiopurity material assays and low background instrumentation development in order to ensure that the challenging sensitivities of the next generation experiments can be achieved on competitive timescales.}

In order to establish a multi-technology and multi-isotope DBD0$\nu$ physics program, extensive underground
space to host the DBD0$\nu$-experiments and related R\&Ds activities is necessary. In Europe the Gran Sasso underground laboratory has the required depth and could host all currently proposed next generation DBD0$\nu$ European experiments. At the same time all  other underground laboratories must be strongly involved in the present DBD0$\nu$ strategy to support various R\&D phases for detector development and to guarantee sufficient resources for material selection and detector design.  Pilot experiments will be needed to implement complex and costly experimental apparatus and  onsite expertise in low-background techniques is necessary for an effective and timely implementation of the experimental programs. In order to pursue the next generation of neutrinoless double beta decay experiments, a close coordination between the European underground laboratories in the areas of low-background instrumentation development, detector prototyping and radiopurity screening is therefore mandatory.

\vspace{2truemm}

{\em Recommendation 6. The theoretical assessment of the particle physics implications of a positive observation and of the broader physics reach of these experiments should be continued. A dedicated theoretical and experimental effort, in collaboration with the nuclear physics community, is needed to achieve a more accurate determination of the Nuclear Matrix Elements (NME).}

Once and if a positive signature is found, lepton number violation will be established and a key question will be to establish the physics mechanism behind DBD$0\nu$, whether it is indeed light Majorana neutrinos or a more exotic one. A strong theoretical effort should be devoted to continue to explore different theoretical models behind neutrinoless double beta decay and the complementarity with other experimental searches. Identifying the LNV mechanism mediating this process will allow to extract additional information on the particles involved. Most interestingly, in the case of the simplest mechanism of light neutrino exchange, the measurement of the half-life would give information on the values of neutrino masses and, at least in principle, on Majorana CP violation, with a strong complementarity with the determination of neutrino masses from cosmology.

Such plans require to extract the effective Majorana mass parameter with high precision, for which nuclear matrix elements need to be evaluated. The computation of NME is challenging and currently is affected by an uncertainty which is typically quantified as a factor of 2-3. New developments are very promising and exploit ab-initio computations. An enhanced theoretical effort is  required. Stronger interactions between the particle and nuclear physics communities would be highly beneficial and a closer coordination with NuPECC is recommended.
Further progress with NME calculations requires experimental input from a range of nuclear process measurements. In particular, detailed studies of $2\nu\beta\beta$ decay differential characteristics (such as single electron energy spectra and angular distributions) help understand the importance of contributions from intermediate nuclear states. Together with muon-nucleus capture experiments, they can help resolve the ``$g_A$ quenching" dilemma, while ion charge exchange reaction experiments provide constraints for nuclear wave functions of the initial and final states of the DBD process.

Thanks to the large mass, low background and high detector performances, the next generation double beta decay experiments will be sensitive also to a certain number of other physics processes,  allowing experimental investigation with unprecedented sensitivities. These include alternative double beta decay modes, some exotic processes predicted by the extensions of the Standard Model, validation of fundamental physics principles and, most importantly, the search for interactions of  Dark Matter particles. The upcoming experiments have therefore a multipurpose nature with neutrinoless double beta decay at its core, but with the added value of exploring other avenues of significant scientific importance.

\newpage

\tableofcontents

\newpage

\section{Introduction}

With the discovery that neutrinos have mass, thanks to neutrino oscillations, a key question, and arguably the most important in neutrino physics, concerns the nature of neutrinos - whether they are Majorana or Dirac particles. This question is intrinsically related to the conservation or not of lepton number and can be best, and in most cases, only addressed by neutrinoless double beta decay (DBD$0\nu$).

Lepton number violation (LNV) is a crucial question in particle physics for several reasons:
\begin{itemize}
    \item Lepton number, as baryon number, is an accidental symmetry in the Standard Model (SM), i.e. it happens to be respected by the SM Lagrangian because of the gauge structure and the SM particle content. It is violated at the non-perturbative level, with only the $B-L$ combination being preserved.  Whether it is a fundamental symmetry of nature or not is a central question in particle physics.
    \item Gravitational effects are expected to induce breaking of global symmetries such as lepton number. Specifically, it has been recently shown that quantum gravity imposes symmetries to be either gauged or broken~\cite{Harlow:2018jwu}.
    \item Symmetries are the guiding principle we use to understand  particle interactions. Knowing whether lepton number is violated or not is essential to build the theory beyond the Standard Model implied by the existence of neutrino masses. Interestingly, the lowest dimension effective term that can be added to the SM is the Weinberg operator $(L \cdot H) (L \cdot H) /\Lambda$, where $L$ is the leptonic doublet, $H$ the Higgs one and $\Lambda$ is the heavy scale at which the full theory is in action. Higher order terms are suppressed by higher powers of the heavy mass and thus it is not surprising that the new physics BSM has manifested itself in terms of neutrino masses. Most models which explain not just the masses but also their smallness invoke LNV, predict that neutrinos are Majorana particles, and would induce neutrinoless double beta decay. The prime example of these is the see-saw type I mechanism.
    \item
    In the leptogenesis mechanism, lepton number (or baryon number), together with C- and CP-, violation is essential to dynamically generate the baryon asymmetry we observe in the Universe~\footnote{In some specific models an effective breaking of lepton number is achieved in the thermal plasma, while lepton number is conserved overall. Models of this kind go under the name of Dirac leptogenesis.}. Observing LNV in neutrinoless double beta decay and CP violation would provide a strong hint in favour of leptogenesis as the origin of the baryon asymmetry of the Universe.
\end{itemize}

Neutrino oscillations conserve lepton number and cannot distinguish between Majorana and Dirac particles. To test this symmetry and establish the nature of neutrinos, it is necessary to search for processes which break lepton number. The most sensitive of these is neutrinoless double beta decay.

This process takes place in nuclei when two neutrons simultaneously decay into two protons and two electrons, with no neutrino emission. Its SM counterpart is the two-neutrino double beta decay (DBD$2\nu$) in which two electron antineutrinos are produced:
\begin{eqnarray}
 \mathcal{N} (A,Z) &\rightarrow \mathcal{N} (A,Z+2) + 2 e^- + 2 \overline{\nu}_e & \quad  \mathrm{for~DBD 2\nu} ~, \\
  \mathcal{N} (A,Z) & \rightarrow  \mathcal{N} (A,Z+2) + 2 e^- \phantom{+ 2 \overline{\nu}_e}& \quad  \mathrm{for~DBD 0\nu} ~. 
\end{eqnarray}
The existence of two-neutrino double beta decay was first proposed by M. Goeppert-Mayer in 1935. After E. Majorana showed that the results of the beta decay theory do not depend on neutrinos being their own antiparticles, i.e. Majorana particles, or not, W. H. Furry suggested that double beta decay could proceed without neutrino emission, i.e. neutrinoless double beta decay.

Differently from the two-neutrino double beta decay, neutrinoless double beta decay violates lepton number by two units and is not allowed by the SM. For this reason, its discovery would be of paramount importance and would imply that neutrinos are of Majorana type, unlike all other fermions in the SM. There is a rich interplay with neutrino oscillation experiments, in particular with regards to the neutrino mass ordering and the question of  CP-violation in the lepton sector, with cosmology, for neutrino masses, as well as with collider experiments carrying out complementary searches for lepton number violation.  

\vspace{5truemm}
Double beta decays can be searched for in nuclei in which single beta decay is kinematically forbidden. Typical nuclei considered are $\mbox{}^{48}$Ca, $\mbox{}^{76}$Ge, $\mbox{}^{100}$Mo, $\mbox{}^{130}$Te, $\mbox{}^{136}$Xe, $\mbox{}^{150}$Nd, among others. The typical signature is the observation of two electrons which present a continuum spectrum for DBD$2\nu$ and a narrow peak at the $Q$-value in the case of DBD$0\nu$, see Fig.~\ref{fig:spectrum}.

\begin{figure}
    \centering
    \includegraphics[width=0.6\textwidth]{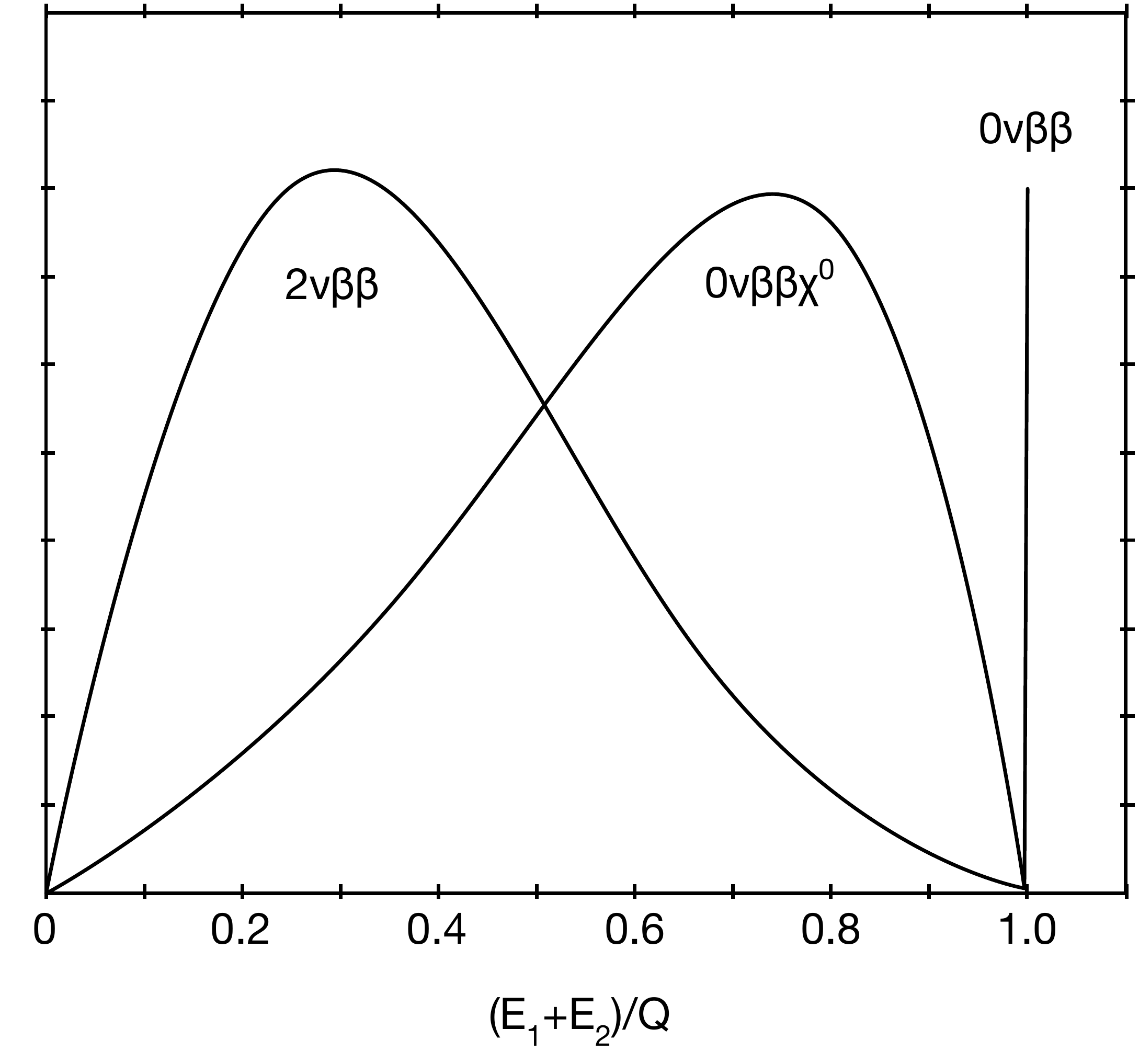}
    \caption{Spectra for the sum of the kinetic energies of the two emitted electrons for two-neutrino double beta decay, neutrinoless DBD and for double beta decay with Majoron emission. The amplitudes are taken as arbitrary.}
    \label{fig:spectrum}
\end{figure}

There is a vibrant and diverse program of DBD$0\nu$ research world-wide with a number of experiments featuring prominently on scientific roadmaps in Europe, North America and Asia. Europe in particular has established a recognised  leadership and an outstanding track record in the field through its most prominent contributions to a number of experiments (CUORE/CUPID, GERDA/LEGEND, NEXT, NEMO-3/SuperNEMO). 

The experimental approaches currently pursued world-wide can be broadly categorised in four main categories: (1) large liquid scintillator detectors (KamLAND-Zen, SNO+); (2) high-energy resolution solid-state devices such as High-Purity Germanium (HPGe) detectors (GERDA, Majorana, LEGEND) and cryogenic particles detectors (CUORE, CUPID, AMORE), normally indicated as bolometers; (3) Time Project Chambers (TPC) with Xenon in liquid and gaseous form (EXO-200, nEXO, NEXT, DARWIN); and (4) tracking-calorimeter detectors with a full reconstruction of the final state topology (SuperNEMO). There are also smaller projects at various stages of development that pursue an R\&D program using novel approaches (e.g. COBRA). 

The past ten years have seen an outstanding progress in ultra-low background technologies that are required for the current and next generation of DBD$0\nu$ experiments. Two-neutrino double beta decay allowed in the SM has been observed in ten isotopes in direct "counting" experiments. Background levels at the order of $10^{-3}$ counts keV$^{-1}$ kg$^{-1}$ yr$^{-1}$ have been reached in the DBD$0\nu$ energy region of interest (e.g. GERDA) and large detectors with tens and even hundreds of kg of $\beta\beta$ isotopes have been employed (e.g. KamLAND-Zen). 

In the absence of observation, the tightest constraints have been achieved with the isotopes of 
$^{136}$Xe (KamLAND-Zen), $^{76}$Ge (GERDA) and $^{130}$Te (CUORE) with lower bounds on the half-lives as high as $10^{26}$ yr and corresponding upper limits on the effective Majorana neutrino mass of (0.06 -- 0.2) eV, depending on the nuclear model involved in extracting the lepton number violating parameter. 

The goal of the next generation DBD$0\nu$ experiments is to completely cover the so-called inverted ordering of  neutrino masses whereby the electron neutrino flavour is carried by the more massive neutrino eigenstates. This gives rise to an effective Majorana neutrino masses in the (15 -- 50) meV range. Importantly, these experiments will have a significant discovery potential even in the case of the normal ordering of neutrino masses. 

Given the scale and cost of future experiments, it is widely recognised that a consolidation of the international effort is required. The aim of this document is the development of a strategy that will secure and enhance European leadership in the next generation of the experiments. We will focus on experimental approaches that build on a significant previous investment from European participants and promise to be the most competitive projects --- CUPID, LEGEND and NEXT. This will be reviewed in the international context with other ambitious techniques pursued worldwide, most notably with nEXO, KamLAND-Zen and SNO+ experiments. We note that due to existing uncertainties in nuclear models that affect the interpretation of the observed signal and significant risks posed by unprecedentedly low background requirements it is vital that several isotopes and experimental techniques are employed to search for the DBD$0\nu$ process. The experimental observation of lepton number violation is one of the most pressing tasks in modern particle physics and warrants a significant investment in this very fertile research area with significant discovery potential.

\section{Theoretical aspects of neutrinoless double beta decay}

Neutrinoless double beta decay arises from lepton number violating physics beyond the Standard Model. As oscillation data imply that neutrinos have mass, if the latter are of Majorana type, they will induce DBD$0\nu$ with rates that could be accessible in current and future experiments. For this reason, the exchange of three light Majorana neutrinos should be considered the simplest and is the most studied mechanism mediating neutrinoless double beta decay. Nevertheless, as neutrino masses require an extension of the Standard Model, generically any model which advocates lepton number violation for neutrino masses will also induce DBD$0\nu$ at some level. In some cases, these mechanisms could even dominate over that of light Majorana neutrinos. If and when a discovery is made, a key question will be to establish the dominant mechanism, whether it is indeed light Majorana neutrinos or a more exotic one. This will consequently allow to extract useful information on the particles involved, for instance in the case of light neutrinos to get information on neutrino masses and possibly on CP violation.

\subsection{The three-light Majorana neutrino exchange}

Neutrino oscillation experiments have established that there are, at least, 3 light massive neutrinos with masses $m_1, m_2, m_3$, and have measured the mass squared differences with best fit values $|\Delta m^2_{31}| \equiv |m_3^2 - m_1^2| \simeq 2.5 \times 10^{-3} \ \mathrm{eV}^2$ and $\Delta m^2_{21} \equiv m_2^2 - m_1^2 \simeq 7.39 \times 10^{-5} \ \mathrm{eV}^2$~\cite{Esteban:2018azc}. This leaves open the possibility of two orderings for the neutrino masses: normal ordering (NO) for $m_1 <m_2<m_3$ and inverted ordering (IO) for $m_3<m_1<m_2$.  The overall mass scale is not yet known: the lightest neutrino could be negligible or as heavy as $~0.1$--$0.3 \ \mathrm{eV}$, the latter option corresponding to neutrino masses very close to each other. The massive neutrinos are related to the flavour ones, $\nu_e, \nu_\mu $ and $\nu_\tau$ via the Pontecorvo-Maki-Nakagawa-Sakata $3\times 3$ unitary matrix $U$. The oscillation experiments have also measured with precision the mixing angles which parameterise it, $\theta_{12} \simeq 34^o$, $\theta_{23} \simeq 50^o$ and $\theta_{13} \simeq 8.6^o$. They have provided the first hints in favour of CP violation due to the $\delta$ phases being different from 0 or $\pi$~\cite{Esteban:2018azc}. The two Majorana phases $\alpha_{31}$ and $\alpha_{21}$ in $U$ are physical if neutrinos are Majorana particles and are completely unknown at present.

\begin{figure}
    \centering
    \includegraphics{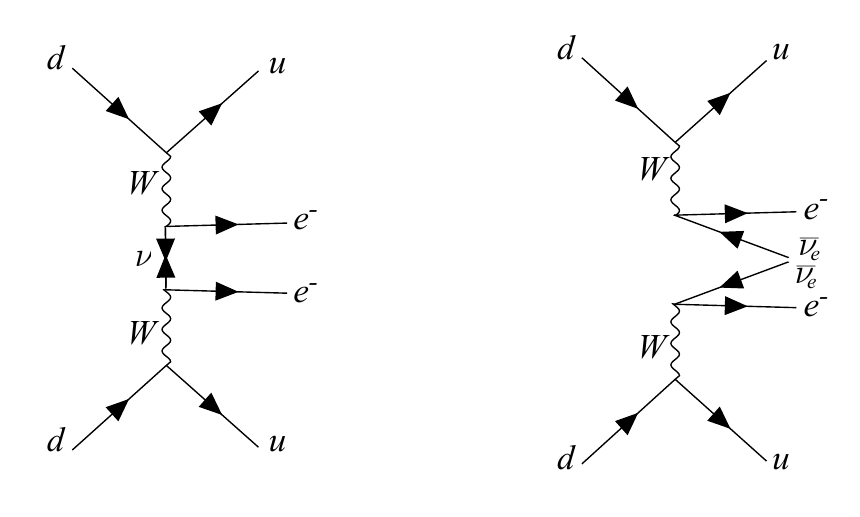}
    \caption{Feynman diagrams at the quark level for neutrinoless double beta decay (left) and for the SM allowed two-neutrino double beta decay (right).}
    \label{fig:Feynman}
\end{figure}

Light Majorana neutrinos mediate neutrinoless double beta decay through the diagram shown in Fig.~\ref{fig:Feynman}.
The half-life is given by 
\begin{equation}
\label{Tdecay}
(T_{1/2}^{0\nu})^{-1} \simeq \frac{G_{0\nu}}{m_e} \,  |m_{\beta \beta}|^2 ~M_{\mathrm{NUCL}}^2~,
\end{equation}
where $G_{0\nu}$ is a known phase-space factor, $m_e$ is the electron mass, $M_{\mathrm{NUCL}}$ is the nuclear matrix element for the nucleus of the process.  $m_{\beta \beta}$ is the  {\it effective Majorana mass parameter} which parameterises all the decay rate dependence on the neutrino quantities, namely the neutrino masses, mixing angles and CP-violating phases.
Restricting the discussion to the standard case of 3-neutrino mixing, its expression is given by
\begin{equation}
|m_{\beta \beta}| \equiv \left| m_1 |U_{\mathrm{e} 1}|^2 
+ m_2 |U_{\mathrm{e} 2}|^2~e^{i\alpha_{21}}
 + m_3 |U_{\mathrm{e} 3}|^2~e^{i(\alpha_{31} - 2 \delta)} \right|~.
\label{effmass2}
\end{equation}
Here, $m_i$, $i=1,2,3$, indicate the three light neutrino masses, which can be expressed in terms of the measured mass squared differences $\Delta m^2_{31}$ and $\Delta m^2_{21}$ and an unknown overall scale set by the lightest neutrino mass, $m_1$ for normal ordering and $m_3$ for inverted ordering.
$U_{ei}$ are the elements of the first row of the PMNS lepton mixing matrix which depend on the angles $\theta_{12}$, $\theta_{13}$ and on the CP violating phases $\alpha_{21}/2$ and $-\delta + \alpha_{31}/2$. The latter phases are unknown and need to be taken as free parameters.

From Eq.~\eqref{effmass2} we see that the predicted value of $m_{\beta \beta}$ depends critically on the 
neutrino mass spectrum and on
the values of the two unknown
Majorana phases
$\alpha_{21}$ and $\alpha_{31}$.
We find that 
\begin{eqnarray}
|m_{\beta \beta}^{\rm NO,m_1\sim 0}  |
 \simeq &
\left|\sqrt{\Delta m^2_{21}} \sin^2\! \theta_{12} \cos^2\! \theta_{13}
                                        + \sqrt{\Delta m^2_{31}} \sin^2 \!\theta_{13}e^{i (\alpha_{32} - 2 \delta) }\right| &
 \simeq  1.1-4.2 \ \mathrm{meV} ,
                                        \label{NH}\\
|m_{\beta \beta}^{\rm IO, m_3\sim0} |
 \simeq & 
 \sqrt{|\Delta m^2_{32}|}  \cos^2\! \theta_{13}\sqrt{ 1 - \sin^2 {2 \theta_{12}} \sin^2 \left( \frac{\alpha_{21}}{2} \right)} &
 \simeq 15 - 50 \ \mathrm{meV} ,
                                        \label{IH}\\
|m_{\beta \beta}^{\rm m_1\simeq m_2\simeq m_3 \equiv m_0} |
\simeq &
m_0 \left|
 (\cos^2 \! \theta_{12}+ \sin^2 \!\theta_{12} e^{i \alpha_{21}} ) \cos^2 \! \theta_{13}
 \! + \! e^{i(\alpha_{31} - 2 \delta)}\! \sin^2 \!\theta_{13} \right| &
\simeq (
0.29 - 1)m_0 ~,
                                        \label{QD}
 \end{eqnarray}
%
where we have used the measured values of the oscillation parameters,  including a $3\sigma$ error, and we varied the CPV phases in their allowed ranges.
In the most general case, we show in Fig.~\ref{fig:spectrum} the current predictions for $m_{\beta \beta}$ for the two mass orderings varying the minimal value of neutrino mass.
\begin{figure}[h]
\begin{center}
\includegraphics[width=0.95\textwidth]{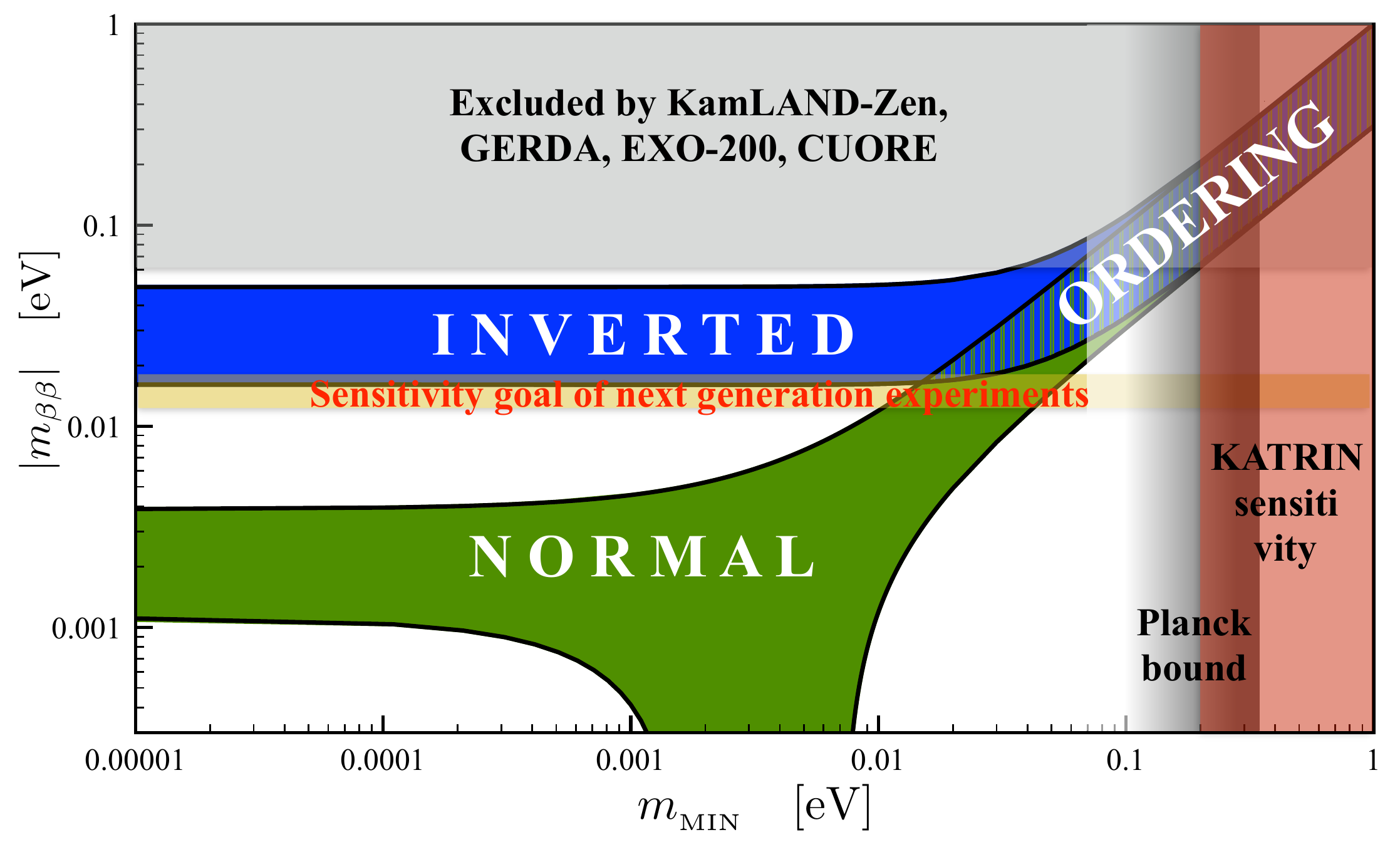}
\caption{The effective Majorana mass $|m_{\beta \beta}|$ as a function of the smallest neutrino mass $m_\mathrm{MIN}$. We have used the current best-fit values and the $2\sigma$ errors of the oscillation parameters from Ref.~\cite{Esteban:2018azc}. 
The Majorana phases $\alpha_{21}$ and $\alpha_{31}$, and $\delta$, are varied within their allowed intervals.}
\label{fig:spectrum}
\end{center}
\end{figure}

As the mixing angle $\theta_{12}$ is large but non-maximal, there is significant lower bound on $m_{\beta \beta}$ for IO given by 
\begin{equation}
|m_{\beta \beta} ^{\mathrm{IO}}|  \geq \sqrt{|\Delta m^2_{32}|} \cos 2 \theta_{12} \simeq 15 ~\mathrm{meV}~.
\label{eqmeffih}
\end{equation}
 In the case of NO the effective Majorana mass can go from the current bound to zero, even if neutrinos are Majorana particles due to a cancellation of the three contributions for values of $m_{\mbox{}_\mathrm{MIN}} \sim 5$~meV, as shown in Fig.~\ref{fig:spectrum}.

Neutrinoless double beta decay can provide information on the neutrino mass spectrum. In the ideal case of perfectly known nuclear matrix elements, a measurement of $ |m_{\beta\beta}| > 0.1$~eV would imply that the spectrum is quasi-degenerate. For values of $|m_{\beta\beta}| < 15$~meV, the ordering would necessarily be normal.
For values in between, both orderings are possible, but with constraints on the masses. For instance, for
$ 15~\mathrm{meV} \leq | m_{\beta\beta} | \leq 50~\mathrm{meV}$, the neutrino mass spectrum
would be either with IO and $m_3<0.2~\mathrm{eV}$ or with NO and partial hierarchy with $m_1 > 15$~meV.
Similar, although somewhat weaker, conclusions can be obtained once the uncertainties on the NME and the experimental error on $m_{\beta\beta} $ are included.

In principle, DBD$0\nu$ could also give information on CP violation due to Majorana phases. A very precise measurement of $m_{\beta\beta} $ together with an accurate determination of the neutrino masses would open this possibility. However, it is extremely challenging as it would require to know the NME with a very small error, at most at the few 10\% level, which at present seems difficult to achieve.

In this discussion, we have treated the neutrino parameters as independent and taken them within their allowed ranges. In particular, the CP violating phases have been allowed to vary without any constraint. Although generically true, this assumption might not be justified in models which aim at understanding the leptonic flavour structure. These models typically invoke some underlying principle, such as a flavour symmetry, which can explain the specific values of the mixing angles that have been observed. In this type of approach, the number of free parameters is greatly reduced, often to just one or two, leading to correlations between the mixing terms.  In models which additionally impose a generalised CP symmetry, it is possible to obtain predictions also for the Majorana CP-violating phases and consequently a much more predictive range of values for $m_{\beta\beta}$. Examples of this type are models based on a discrete symmetry such as $A_4, S_4, A_5$ or on flavour $U(1)$. For instance, if the neutrino mass matrix is invariant under a $\mu-\tau$ reflection, one can show that the mixing angle $\theta_{23}$ and Dirac CP phase $\delta$ are maximal and the Majorana CPV phases are trivial. Other conclusions can be drawn for different symmetries.

\subsection{Other mechanisms for DBD$\nu$}
In presence of lepton number violation in extensions of the Standard Model, it is generically expected that a contribution to DBD$0\nu$ is induced. Models can be separated in two classes depending if the particles mediating the process are heavier or lighter than the typical momentum exchange $\sim {\cal O}( 100 \ \mathrm{MeV})$, leading to short or  long range processes. Light Majorana neutrinos belong to the second class. Mediators, such as heavy sterile neutrinos with $M\gg (100 \ \mathrm{MeV})$, will lead to a suppression by a heavy scale in the propagator and will typically, but not always, be subdominant.

We discuss here the most studied cases:
\begin{itemize}
    \item Light sterile neutrinos. A minimal extension of the Standard Model invokes the existence of new neutral fermions, singlets with respect to the SM gauge symmetries. For this reason they are called ``sterile neutrinos". There is no strong theoretical guidance on the scale of their masses which can go from sub-eV region, preferred on the basis of naturalness arguments, to the GUT scale, where sterile neutrinos can arise naturally, e.g. in $SO(10)$. Some indications in favour of eV sterile neutrinos have been found by short baseline oscillation searches, namely LSND and MiniBooNE and some reactor experiments, but they are in tension with disappearance experiments and most notably recent results from IceCube and MINOS+, as well as with cosmology. Nevertheless, the possibility of the existence of sterile neutrinos with masses in the eV to 100~MeV range cannot be discarded.  Indeed, they are present in many models advocated e.g. for dark matter, leptogenesis etc. If they are of Majorana type, they would contribute to DBD$0\nu$ in the same manner as light neutrinos, so that the effective Majorana mass now reads 
\begin{equation}
m_{\beta \beta}^{n\nu} \equiv  m_{\beta \beta}^{3\nu} + \sum_j m_j U_{\mathrm{e} j}^2 
 ~,
\label{effmass21}
\end{equation}
where, $m_{\beta \beta}^{3\nu}$ is the contribution from the three light standard neutrinos, $m_j$ is the mass of the light nearly-sterile neutrinos and $U_{ej}$ is their mixing parameter with electron neutrinos.
    
    Depending on the values of the masses and mixing angles, they could even give a dominant contribution to neutrinoless double beta decay so that a larger value of $m_{\beta \beta}$ than in the standard case could be found. Partial cancellations could also be present reducing the predicted value of $m_{\beta \beta}$.
    A special case arises if these sterile neutrinos are at the origin of neutrinos masses in a light see-saw mechanism. If their masses are all below the 100~MeV scale, their contribution to $m_{\beta \beta}$ would exactly cancel out that of the three light standard neutrinos, leading to no neutrinoless double beta decay. For this to happen the mixing angles between the heavy neutrinos and the active ones would need to be sizable and it would be possible to search for them in other ways, e.g. oscillations at short baseline, kinks in the beta decay spectrum, searches of additional peaks in the electron spectrum in pion and kaon decays and others.
    \item
    \begin{figure}[t]
        \centering
        \includegraphics[width=0.9\textwidth]{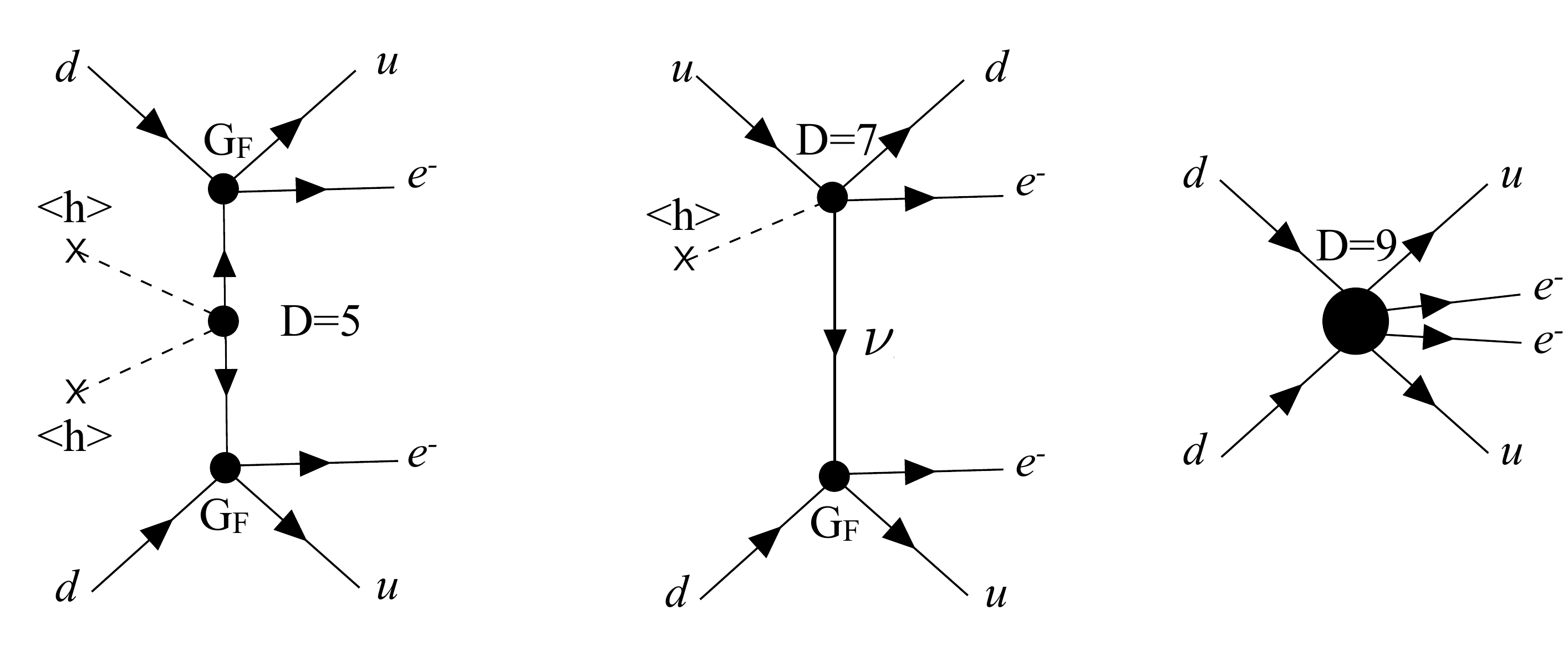}
        \caption{Contributions to neutrinoless double beta decay  from effective LNV operators of higher dimension: $D=5$ Weinberg operator which corresponds to the standard light mass mechanism (left) , $D=7$ operator inducing long–range
contributions (centre), $D=9$ operator leading to short–range contribution (left).}
        \label{fig:othermec}
    \end{figure}
    Effective Lagrangian. Without the need to specify the exact nature of the lepton number violating physics, it is possible to summarise its effects at low energy in terms of higher order operators which are added to the Standard Model Lagrangian. The ones relevant for DBD$0\nu$ are depicted in Fig.~\ref{fig:othermec}.
    The first one is the $D=5$ operator responsible for neutrino masses. The second one corresponds to $D=7$ long range interactions and can give a sizable contribution to neutrinoless double beta decay.
    The third one is a $D=9$ operator, which for instance can be generated in the presence of heavy nearly-sterile neutrinos. As we notice these terms are suppressed by a new heavy scale and the naive expectation is that these terms will be subdominant with respect to the light Majorana mass mediation, unless the scale of the new physics is not too heavy. We should also point out that the $D=7$ and $D=9$ operators can also induce neutrino masses, leading to very strong constraints on the couplings and on the new physics scale. As a result, typically, the light neutrino mass mechanism dominates unless neutrino masses arise only at the 2- or 3-loop level. 
    \item Heavy sterile neutrinos. They will induce a $D=9$ operator. If the mass scale of the nearly sterile neutrinos is heavier than the momentum exchange, their contribution is typically suppressed as $U_{eN}^2 /M_N$. Nevertheless if they are not much heavier than the GeV scale, they can give a sizable contribution to the process and conversely neutrinoless double beta decay can put significant constraints on their masses and mixing with electron neutrinos. An independent test of their existence can be obtained in peak and decay searches, leading to an interesting complementarity in establishing their properties, in case a positive signal is found.
    \item
    Left-right symmetric models. Interesting extensions of the Standard Model introduce a right chiral sector at higher energies, in parallel to the Standard Model left-handed one. These models include right-handed vector bosons, right-handed neutrinos and a new scalar sector necessary to break the $SU(2)_R$ symmetry to the SM one. These terms result in $D=9$ operators which are not suppressed by the small heavy-active mixing angles but by a right-handed vector boson mass, constrained to be above the TeV scale, to the 4th power. Other diagrams involve mixing with the active neutrinos or the exchange of scalar triplets. An interesting synergy is present with collider experiments which can test the existence of these new gauge bosons, of the new scalars and of the heavy sterile neutrinos. 
\item
Supersymmetry. In supersymmetric models, R parity violation allows terms involving one lepton and two quarks, which violate lepton number. Neutrinoless double beta decay can proceed via the exchange of supersymmetric particles both at short range and long range, in which case the supersymmetric particles are involved in just one vertex. Strong constraints on the relevant couplings $\lambda^\prime_{idk}$ can be derived for masses in the TeV range.
\item
Leptoquarks. These are scalars or vector bosons which couple both to leptons and quarks and emerge in GUTs, extended technicolor and/or composite models. In presence of lepton number violation, they can mediate DBD$0\nu$ via long range interactions, e.g. for leptoquark-Higgs coupling.
\item
Extra dimensions. Models with extra dimensions have been invoked to solve the Standard Model naturalness problem. They can have towers of Kaluza-Klein states which can mediate DBD$0\nu$-decay if lepton number is violated. Interestingly, these towers can have states with masses both below and above the typical momentum exchange and can avoid the connection between neutrino masses and neutrinoless double beta decay.
\end{itemize}

All in all, there are many mechanisms which can induce neutrinoless double beta decay and, if a signal is found, it will be of paramount importance to identify the dominant contribution, testing if the standard light neutrino mass exchange is indeed the most important one. Generically, it is expected that short-range interactions are subdominant, unless the new mass scale is not too heavy. Long range processes, due to light neutrino masses and other exotic physics, could be at play.
The complementarity with other new physics search is essential in this endeavour.

It is also important to identify observables which can distinguish between short range and long range processes and different mediators. It has been pointed out that the NME for these processes can scale differently with the type of nuclei. In principle, measuring the decay rates in different nuclei would allow to disentangle the two types of contributions. Particularly advantageous combinations that have been identified are e.g. $\mbox{}^{76}$Ge vs $\mbox{}^{136}$Xe and $\mbox{}^{100}$Mo vs $\mbox{}^{136}$Xe.
The angular distribution between the two electrons is also very important but a dedicated effort, as the one proposed in SuperNEMO, would be required to measure it.

\subsection{Complementarity with other searches}

Neutrinoless double beta decay has a unique role in testing LNV. Its complementarity with other searches greatly enhances the physics reach and could allow to answer questions that each individual approach cannot address by itself. 

\vspace{0.1truecm}

{\bf Neutrinoless double beta decay and neutrino experiments.} The predictions for $m_{\beta \beta}$ depend on the neutrino mass ordering. If neutrino oscillation experiments determine that the neutrino mass ordering is inverted, $|m_{\beta \beta}|$ is predicted to be bigger than 15 meV providing a clear target for neutrinoless double beta decay experiments. Further conclusions could be obtained depending on the experimental results. We give here some relevant examples. 

Let’s first assume that the ordering is established to be inverted in long-baseline neutrino experiments. (i) If $|m_{\beta \beta}| \geq 15$~meV, neutrinos are Majorana particles. Moreover, if $|m_{\beta \beta}|> 50$~meV both upper and lower bounds on $m_3$ can be deduced, given approximately by $|m_{\beta \beta}|  \leq m_3 \leq |m_{\beta \beta}|/ \cos 2 \theta_{12}$. Consequently, a predicted range for the sum of neutrino masses relevant in cosmology could be found and could be confronted with observations. For $15 \ \mathrm{meV} \leq |m_{\beta \beta}| \leq 50$~meV, the spectrum would be inverted hierarchical (IO and $m_3\sim0$). (ii) If $|m_{\beta \beta}| < 15$~meV is measured, neutrinos are also established to be Majorana particles but a cancellation between the standard light neutrino mass contribution and new physics is necessary. 
(iii) If only an upper bound below 15 meV is found on $|m_{\beta \beta}|$, then the simplest conclusion is that neutrinos are Dirac particles, although with a caveat that a cancellation between the three-neutrino contribution and new physics could still be at work, for instance in the case of a light see-saw. It would be crucial to test this second possibility by looking for new particles and interactions which can be responsible for this cancellation. 

Let’s now consider the case in which neutrino oscillation experiments determine that the ordering is normal, as initial current hints seem to indicate. The predictions for $m_{\beta \beta}$ go from the current bounds to a complete cancellation (see Fig. 1) if neutrino masses have a partial hierarchy. A measurement of $m_{\beta \beta}$ would establish that neutrinos are Majorana particles and would restrict their masses to a specific range in a similar manner as discussed above. 

A strong complementarity is also present with direct measurement of neutrino masses, in beta decay experiments. KATRIN is starting to test values of neutrino masses below the eV scale, with an ultimate sensitivity to 0.2~eV. Improving on this will be challenging but new experimental strategies are being explored in e.g. Project 8. These experiments would provide a model-independent measurement of neutrino masses, which would have to be confronted with that of $m_{\beta \beta}$ to possibly get information on Majorana CP violation and/or to identify incompatibilities pointing towards new mechanisms behind neutrinoless double beta decay. It is worth to point out that if KATRIN obtains a positive signal, and no signature is found in DBD$0\nu$ so far, one should conclude that neutrinos are Dirac particles or should hunt for possible sources of partial cancellation of the 3-light neutrino contribution. 

\vspace{0.1truecm}

{\bf Neutrinoless double beta decay and cosmology.}
Cosmological observations are providing the most sensitive test of neutrino masses, albeit with strong underlying assumptions concerning the cosmological model and the thermal distribution of relic neutrinos. Current studies seem to indicate that they will be able to distinguish between normal (NO with $ m_1\sim 0$) and inverted hierarchical (IO with $ m_3\sim 0$) spectra with implications for the predictions in neutrinoless double beta decay.

\begin{figure}[h]
\centerline{
\includegraphics[width=10cm]{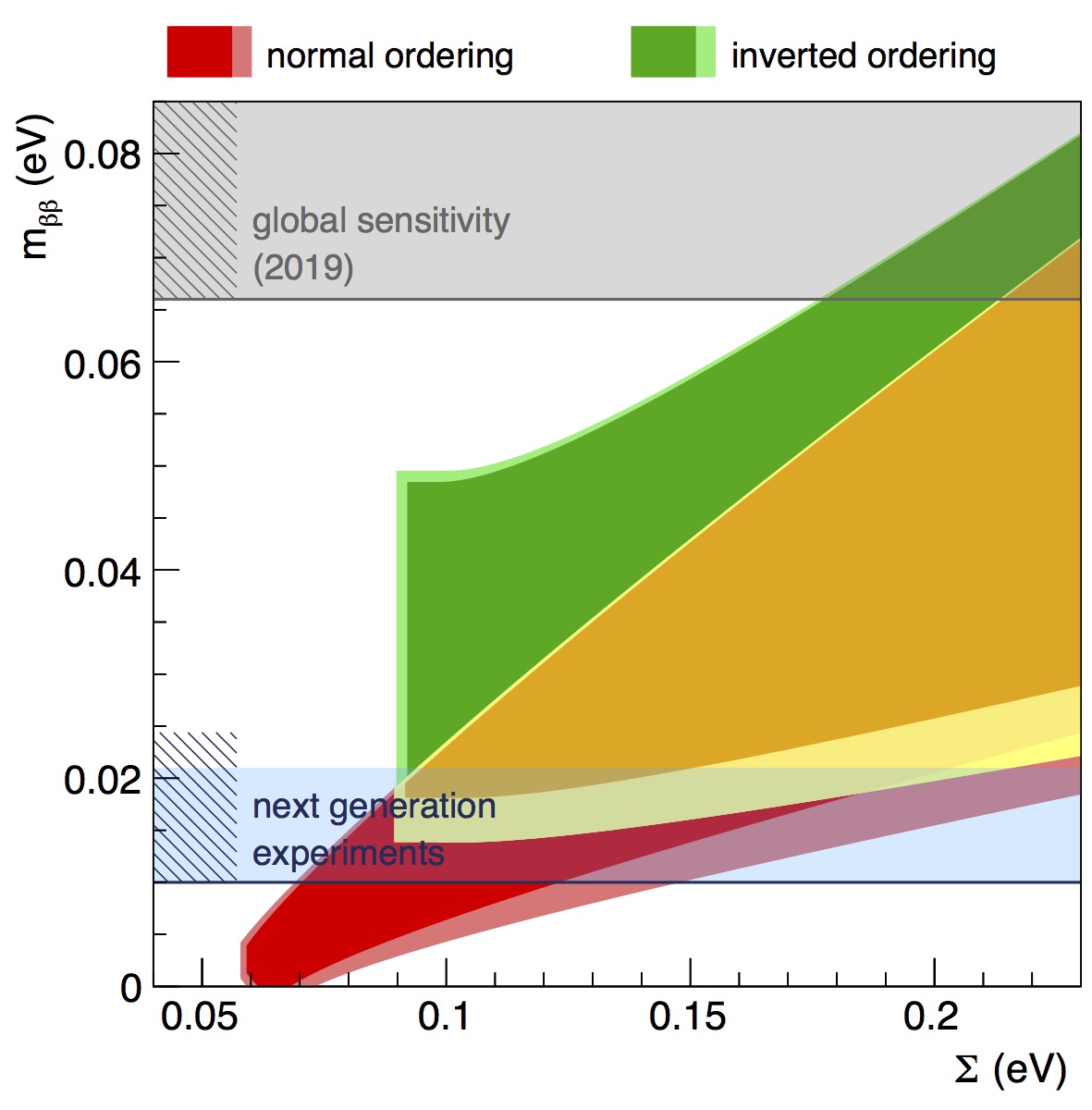}}
\caption{The effective Majorana mass $m_{\beta \beta}$ as a function of the sum of neutrino masses $\Sigma_i m_i\equiv \Sigma$. 
The Majorana phases $\alpha_{21}$ and $\alpha_{31}$, and $\delta$, are varied within their allowed intervals $[0,180^\circ]$. The horizontal lines indicate the combined sensitivity of current experiments and the prospective ones for the next generation. Figure from \cite{Wiesinger}.}
\label{fig:meffvscosmo}
\end{figure}

In Fig.~\ref{fig:meffvscosmo} we show the predicted values of $m_{\beta\beta}$ versus the sum of neutrino masses measurable by cosmological observations.

In principle, the combination of a precise measurement of the masses from cosmological observations and of $m_{\beta \beta}$ may allow to get information on CP violation due to Majorana phases. This is a very challenging search, which requires a significant improvement in the evaluation of the nuclear matrix elements.

We also point out that an incompatibility with a possible measurement of neutrino masses from the two approaches could point towards new physics beyond 3  light neutrinos. For instance, if a measurement of the half life indicates a too large value of masses compared to cosmological observations, this may be due to additional contributions to neutrinoless double beta decay or a revision of the cosmological model may be necessary. 

\vspace{0.1truecm}

{\bf The neutrinoless double beta decay-collider interplay.} Models leading to DBD$0\nu$ at short range involve new heavy particles that can be produced directly at colliders, such as the LHC, and lead to lepton number violating signatures, e.g. same sign dileptons and jets. Typically, current LHC bounds are more stringent than those from neutrinoless double beta decay due to the fact that in the former particles are produced on resonance, while in the latter they lead to a strong suppression. Except for few special cases, bounds from the LHC imply that these contributions are subdominant in DBD$0\nu$. Conversely, if neutrinoless double beta decay is mediated by short-range interactions it is generically expected that LNV signals will be seen in present and future colliders and in other LNV searches, e.g. meson and tau decays.

\vspace{0.1truecm}

{\bf Neutrinoless double beta decay and leptogenesis.} Both the generation of the matter-antimatter asymmetry via standard leptogenesis~\cite{Fukugita:1986hr} and neutrinoless double beta decay require the violation of lepton number~\footnote{In Dirac leptogenesis it is possible to achieve leptogenesis with Dirac neutrinos and with lepton number conservation. An equal and opposite lepton asymmetry is generated in a left-handed and a right-handed neutrino sectors that communicate very weakly in the Early Universe. As sphaleron effects affect only the left-handed one, a net baryon asymmetry is generated~\cite{Dick:1999je,Murayama:2002je}. Therefore, the non-observation of lepton number violation would not disproof leptogenesis as a mechanism for the baryon asymmetry of the Universe.}. It is therefore possible to ask if the two are related. The simplest example is that of see-saw type I models at scales much higher than the electroweak one, which assume the existence of heavy sterile neutrinos. In such models, light neutrino masses come from a $D=5$ Weinberg operator arising from the exchange of the heavy sterile neutrinos. In the early Universe, these new fermions were in thermal equilibrium and subsequently decay once the temperature drops below their mass. In presence of lepton number and of C and CP violation, their decays generate a lepton asymmetry which is then converted into a baryon asymmetry by Standard Model non perturbative effects, called sphalerons. Recently, it has been shown that generically low energy CP violation, including that due to the Majorana phases, induce a baryon asymmetry that could be even compatible with the observed values.

Leptogenesis could also be due to other mechanisms involving lepton number violation. If the latter are observed at the LHC, these same processes together with the Standard Model $B+L$ sphalerons  wash out any pre-existing baryon asymmetry, indicating that its origin has to be at relatively low scale. If DBD$0\nu$ is observed there are two possibilities:
\begin{itemize}
    \item this process is mediated at long range, typically by the light neutrino masses, whose origin relies on new physics at a high energy scale, e.g. a see-saw mechanism, responsible also for leptogenesis. An alternative is that both neutrino masses and the baryon asymmetry come from GeV scale see-saw models, that are testable in peak searches and beam dump experiments.
    \item if it is due to short range interactions, LNV signatures would be expected at the LHC, and a mechanism at the TeV-scale or below for baryogenesis needs to be advocated.
\end{itemize}
These generic statements have some loopholes, if for instance lepton number is not violated universally but only in some flavour, if some new hidden symmetry can be advocated to protect the baryon asymmetry from LNV washout or if LNV emerges below the electroweak scale when sphaleron effects are no longer active.

\vspace{0.1truecm}

{\bf Neutrinoless double beta decay and proton decay.} Keeping in mind the special nature of $B-L$ in the SM and beyond, it is noteworthy that searches for lepton number violation are accompanied by those for baryon number violation, in particular proton decay and neutron-antineutron oscillation~\cite{Dolinski:2019nrj}. These processes test different energy scales with proton decay reaching $10^{16}$ GeV and have a strong complementarity in searching for new physics beyond the SM.

\section{Nuclear matrix elements for neutrinoless double beta decay}

Nuclear matrix elements for DBD$0\nu$ are essential to obtain predictions for the half-life of this process given the current knowledge of neutrino masses and, even more so, once and if a positive signal is found, in order to extract the value of $m_{\beta \beta}$ or equivalent parameters for exotic models. NMEs need to be obtained from nuclear theory calculations and require a good description of the initial and final nuclei in the $\beta\beta$ decay (obtained with a nuclear many-body approach), and the evaluation of the DBD$0\nu$ transition operator between these states~\cite{Engel:2016xgb}.

A significant progress has been achieved in their evaluation in the last decade. Nevertheless, there is still a spread by the factor
2-3 between the calculations using different nuclear models. While earlier the evaluations
of NMEs were performed mostly within the Quasiparticle Random Phase Approximation (QRPA)
and nuclear shell model (NSM), nowadays results of the interacting boson model (IBM), and of
different versions of the energy density functional (EDF), are also available. It is generally accepted that
all these models suffer from neglecting certain essential aspects of physics, different in each case. 
However, it is difficult, or impossible, to reliably assign the corresponding uncertainties in the resulting NMEs.
Approaches with ``controlled errors", like no core shell model, coupled cluster methods,
or Green's function Monte-Carlo are being developed. They are, however, so far applicable only to the
light nuclear systems and not yet to the relatively heavy DBD$0\nu$ decay candidate nuclei.

For them, and the various nuclear model based methods, the concrete issues that are widely discussed
are the role of ground state correlations, deformation, the size of the model space, the
restoration of the SU(4) spin-isospin symmetry. The problem of tha so-called quenching of the axial weak
current, often simplified by the concept of ``$g_A$ quenching", is of particular importance. There is not
a consensus on its origin, but some studies indicate that a careful treatment of nuclear correlations
and inclusion of the three-body interaction and of the corresponding two-body weak currents avoids the
need for quenching.

It is expected that the ``quenching" and other limitations of current NMEs calculations will be overcome in the near or mid-term future, once ab initio many-body calculations using nuclear interactions based on QCD can be extended to compute DBD$0\nu$ NMEs. Significant steps have been taken in this direction with very promising results in very recent times.


\subsection{Present NME calculations}
The most reliable NME results are provided by the nuclear shell model, quasiparticle random-phase approximation, energy-density functional theory and interaction boson model. 

\begin{itemize}
    \item 
Nuclear Shell Model (NSM, also called ISM), see e.g. Ref.s \cite{Menendez:2008jp,Caurier:2004gf}. 
The nuclear shell model is a broadly used many-body method in nuclear physics, which can well describe the properties of medium-mass and selected heavy nuclei. It builds on the idea that only the nucleons near the Fermi level are relevant, limiting the configuration or valence space to a subset of the nucleons while the others are frozen in the lowest energy orbitals. The correlations within the space are included and treated carefully using an effective nuclear interaction acting on the configuration space but corrected to account for the effects of the inert core. The nuclear shell model can describe ground-state properties such as masses, separation energies, and charge radii quite well. 
\item
Quasiparticle Random Phase Approximation (QRPA), see e.g. Ref.s \cite{Simkovic:2013qiy,Hyvarinen:2015bda}.
The QRPA considers a large configuration space with a large number of single-particle orbits. The drawback is that the method allows to include only a limited number of correlations and requires a significant modification of the effective nucleon-nucleon interactions used to generate the nuclear states. The proton-proton and neutron-neutron pairings are taken into account and the proton-neutron pairing turns out to be most important. The latter is usually fixed to reproduce the 2-neutrino double beta decay rate with a tuned value for the strength of the coupling. Several modification of QRPA have been proposed to address this problem. 
\item
Energy Density Functional Method (EDF)~\cite{Rodriguez:2010mn}. 
EDF theory focuses on the minimisation of an energy functional 
with respect to quantities such as the number density, the spin density, the current density. In the applications to DBD$0\nu$ decay, the initial and final states are
obtained using the Gogny D1S or a relativistic functional. It can well describe collective properties of the nuclei, for example it can include an explicit calculation of the NMEs as a function of the quadrupole deformation of initial and final nuclei. 
\item
Interacting Boson Model (IBM-2)~\cite{Barea:2015kwa}.
The IBM method tries to merge the advantages of the NSM and EDF methods, namely the inclusion of all correlations of the nucleons around the Fermi surface and the careful treatment of collective motion. It can describe the excitation spectra and the electromagnetic transitions among collective states up to heavy nuclei. It models the low-lying states of the nucleus in terms of bosons, either as s bosons or d bosons, which interact through one- and two-body forces giving rise to bosonic wave functions. 
\end{itemize}

These approaches somewhat disagree in their prediction of NMEs for any DBD emitter, with NSM giving the smallest results, EDF theory the largest, and QRPA and IBM lay somewhere in between~\cite{Engel:2016xgb}.

In general, the NSM and IBM use relatively small configuration spaces for protons and neutrons, while QRPA and EDF theory neglect some nuclear correlations in the large configuration space where they operate. In contrast, the NSM includes all possible correlations within its configuration space~\cite{Engel:2016xgb}.
The NSM and EDF theory are typically the preferred methods in nuclear structure. On the other hand, the QRPA is especially tailored to deal with $\beta$ and $\beta\beta$ decays. 

The strengths and weaknesses of each of the above mentioned nuclear many-body approaches are known~\cite{Engel:2016xgb}. Improved NSM calculations in extended configuration spaces, suggested to enhance NME values, have been performed recently however with little changes on NMEs. In contrast, recent QRPA calculations including additional correlations that allowed deformation led to smaller NMEs, closer to the ones of NSM~\cite{Fang:2018tui}. On the other hand, it has been suggested that EDF theory and IBM NMEs could be reduced if isoscalar pairing correlations could be included explicitly. This is, however, beyond current computational capabilities. 

\subsection{The issue of ``g$_A$ quenching": ab initio NME calculations}

A key benchmark of DBD$0\nu$ calculations is single $\beta$ decay, observed in most unstable nuclei. NSM calculations give too large $\beta$ decay matrix elements, which require a ``quenching" of about $20\%-30\%$. Since matrix elements, at leading order, are proportional to $g_A$, the overestimation is known as ``$g_A$ quenching"~\cite{MartinezPinedo:1996vz}.

The ``quenching" reflects a deficiency of the nuclear theory calculations. It could be caused by limited nuclear correlations in the initial and final nuclei -- pointing to a limitation of the nuclear many-body method -- or alternatively by a simplified transition operator, without meson-exchange currents. This is a matter of concern because the ``quenching" could impact DBD$0\nu$  NMEs. Nevertheless, the $\beta$ and DBD$0\nu$ operators and momentum transfers are quite different, and it is unclear whether DBD$0\nu$ NMEs require a correction similar to the one in single $\beta$ decay~\cite{Engel:2016xgb}.

A very recent work addresses this issue~\cite{Gysbers:2019uyb} using ab initio many-body methods where all nucleons are considered in the calculation, without uncontrolled approximations. Also, the nuclear interactions used are connected to QCD. These are the most advanced nuclear theory calculations, which in the last decade have extended from light to medium-mass nuclei. Energies of ground and excited states are very well reproduced by nuclear ab initio calculations~\cite{Hebeler:2015hla,Stroberg:2019mxo}.

Ab initio $\beta$ decay matrix elements reproduce experimental data without any ``quenching"~\cite{Gysbers:2019uyb}. The analysis of ab initio results suggest that the disagreement in previous calculations is caused, on equal parts, by missing nuclear correlations and meson-exchange currents. This indicates that current NSM DBD$0\nu$ NMEs may not need much ``quenching", because meson-exchange currents are more important in $\beta$ decay~\cite{Wang:2018htk}.

Work is in progress towards extending ab initio calculations to DBD$0\nu$ decay. First matrix elements on the lightest DBD emitters are expected to be completed this year.

\subsection{Experiments connected to neutrinoless $\beta\beta$ decay}

Further progress could be achieved on the evaluation of NMEs and their reliability, if related nuclear processes are going to be pursued both theoretically
and experimentally. One example are matrix elements and differential characteristics of the
two neutrino double beta decay  that share initial and final states with DBD$0\nu$-decay.
DBD$2\nu$ NMEs require the evaluation of the states in the
intermediate odd-odd nucleus, possible so far only in the QRPA and NSM. In this context the role of the
states in the odd-odd nucleus relatively far from the ground state, i.e. near the region of the giant Gamow-Teller
resonance, is crucial. NSM and QRPA have well reproduced DBD$2\nu$ decay lifetimes and they predicted within uncertainties the recently measured two-neutrino double electron capture decay of $^{124}$Xe~\cite{XENON:2019dti}.

Another important input can come from muon capture, a weak process with $\sim$100 MeV of momentum
transfer, where many multipoles play an important role. Experimental work on this process is conducted at J-PARC.
Analysis of the results, as well as of the total muon capture rate, would help in resolving the ``$g_A$ quenching"
dilemma. Relevant information can be provided by the study of heavy-ion double-charge exchange reaction (with $\Delta T = 2$),
in particular of the ground state to ground state transitions, at LNC Catania (the NUMEN experiment). The corresponding cross-sections, and analogous low energy pion double charge exchange reaction, have been suggested to be correlated with DBD$0\nu$ NMEs~\cite{Shimizu:2017qcy}. Experiments in progress at INFN-LNS and RIKEN will measure double charge-exchange reactions in DBD nuclei.
Besides $\beta$ decay, other observables can test the nuclear many-body methods used to obtain NMEs.

Stronger collaboration between the particle and nuclear physics communities involved in neutrinoless double beta decay, and the nuclear experimental one should be encouraged on the topics of common interest discussed above.

\section{Experimental Status of the Art and Future Prospects}
\label{Status and Prospects}

Experimental searches for DBD0$\nu$-decay have a long history with the first experiments dating back to the mid nineties. Apart from aiming for a discovery of the DBD0$\nu$-decay, present and next-generation experiments in particular are focusing on the  exciting and obvious goal to explore / exclude the inverted ordering region. 
Successful experiments in this field have to satisfy common requirements in order to have sensitivity for DBD0$\nu$-decay: \begin{itemize}
    \item an ultra low background;
    \item an excellent energy resolution;
    \item and a large isotope mass.
\end{itemize}   
An optimization of each of these features is specific to each single detector technology.

Experimental approaches differ in the employed isotope, with the most favourable candidates being $^{48}$Ca, $^{76}$Ge, $^{82}$Se, $^{100}$Mo, $^{116}$Cd, $^{124}$Sn, $^{130}$Te, $^{136}$Xe and $^{150}$Nd. The choice of the isotope underlays considerations like the Q-value (possibly above the background $\gamma$ lines), the natural isotopic abundance and the possibility for enrichment as well as compatibility with the targeted technology.  

In Europe in particular three experimental approaches have developed a leading status in both technological development and scientific expertise: 
\begin{itemize}
    \item solid state germanium diodes providing excellent energy resolution combined with event-discrimination between single-side events and multi-side events;
    \item cryogenic scintillating bolometers offering, besides their excellent energy resolution and particle-identification capabilities, also the possibility for employing different isotopes -- of particular interest in case of positive evidence;
    \item gaseous Xe time projection chambers which recently demonstrated good energy resolution and feature, as the only technology thanks to their capability to image the electron tracks, a topological signature.
\end{itemize}

Besides the European experimental projects, an overview is given on other leading technologies in the international context also giving emphasis to large-scale loaded liquid scintillator detectors which, with KamLAND-Zen, currently hold the most sensitive constraint on the effective Majorana mass parameter. 

Since  experiments on DBD0$\nu$ work on a question at the fore-front of research a multi-experimental strategy in both experimental technology and applied isotope is necessary to be pursued as only complementary approaches can finally corroborate the results and findings of each of the individual experiments. 

\subsection{High Purity Germanium Detectors}
Germanium can be enriched in the isotope $^{76}$Ge to a fraction above 87\% and transformed into high-purity germanium (HPGe) detectors. The enrichment does not only increas the signal strength per unit detector mass but also reduces significantly the light isotope fraction and therefore cosmogenic produced backgrounds. Given the intrinsic radio-purity of HPGe detectors, their excellent energy resolution and high signal acceptance for DBD$0\nu$~decays, enriched HPGe detectors have a long-standing record in DBD$0\nu$~ searches. 

\subsubsection{State-of-the-art in $^{76}$Ge  DBD$0\nu$ search}
While the early HPGe experiments used traditional semi-coaxial detectors, the state-of-the-art experiments \Gerda~ \cite{Agostini:2017hit} and \MJD~ \cite{Alvis:2019sil} developed and deployed novel types of HPGe detectors that provide detailed information about the topology of events through the time structure of the recorded charge signal.  Furthermore,  the energy resolution could be further improved by lowering the detector capacitance. Candidate DBD$0\nu$ events can be recognized and discriminated efficiently from spurious interfering signals, as e.g. $\alpha$-events on the detector surface on an event-by-event basis. Also \Gerda~ pioneered the operation of bare HPGe detectors in a high-purity instrumented liquid argon (LAr) shield. LAr provides not only the cooling for the HPGe detectors and the shielding against external radiation but also serves as an active veto system by identifying background events that deposit energy in the LAr. This is of utmost importance for radio-impurities from components close-by to the HPGe detectors as cables, holders or electronic components. Based on these novel concepts, the \Gerda~ experiment,  located at the underground Laboratori Nazionali del Gran Sasso (LNGS) of INFN, Italy, is currently the experiment with the lowest background in the signal region. With a background index (BI) of $5.6^{+3.4}_{-2.4} \cdot 10^{-4}$\ctsper~ and an energy resolution of 3.0~keV (FWHM), or 1.3~keV ($\sigma$)  at \Qbb $= 2039$~keV,  it is the first DBD$0\nu$ experiment which operates free of backgrounds (ie. $<1$ expected event) in the signal window (1~FWHM) within the design exposure.  With a total published exposure of 82.4 \kgyr ,  the derived confidence interval corresponds to $T_{1/2} > 0.9\cdot 10^{26}$~yr (90\% C.L.) to be compared to the median sensitivity of $T_{1/2} > 1.1\cdot 10^{26}$~yr (90\% C.L.) 
\cite{Agostini:2019hzm}
{\sc Gerda}  is thus the first experiment to surmount $10^{26}$~yr sensitivity.

The \MJD~ experiment operates HPGe detectors in a high-purity copper shield, which has been produced in an electro-forming process deep underground. The front-end electronics developed for the \MJD~ achieved an outstanding energy resolution of 2.5~keV (FWHM) at \Qbb . Pulse shape discrimination methods and properties are similar to those of \Gerda, with minor differences given the specific detector architecture of their so-called point-contact detectors and the larger open detector surface area. With an 
exposure 26.0~\kgyr~ \MJD~ achieved a median sensitivity of $T_{1/2} > 0.48\cdot 10^{26}$~yr (90\% C.L.) and a limit of $T_{1/2} > 0.27\cdot 10^{26}$~yr (90\% C.L.) \cite{Alvis:2019sil}.    

\subsubsection{{\sc Legend}}

The \LEG~collaboration aims to develop a phased $^{76}$Ge based double-beta decay experimental program with discovery potential at a half-life beyond $10^{28}$~years, using existing resources as appropriate to expedite physics results \cite{Abgrall:2017syy}.  \LEG~ builds on the successful \Gerda~ and \MJD~ experiments and the collaboration consists of about 50 institutions, about 250 scientists and unites all interested groups in $^{76}$Ge DBD$0\nu$  world-wide. \Lthou~ is amongst the three experiments considered by the US DOE for the ton-scale experiments.

\LEG~ pursues a phased approach incrementing the enriched germanium mass in 200-300 kg steps at a time. This allows to carry out competitive DBD$0\nu$ search while continuously improving the sensitivity. The existing \Gerda~ infrastructure will be used for the first stage, named \Ltwo~ to obtain near-term physics results. \Ltwo~  is largely funded and  physics data taking is scheduled to commence in 2021. About 200~kg of enriched germanium detectors will be operated in \Ltwo~ to achieve an exposure of 1000~\kgyr~ with a discovery potential of  $10^{27}$yrs.  The background in \Ltwo~ will be reduced to $<2 \cdot 10^{-4}$ \ctsper~ which corresponds to a factor of 3 with respect to \Gerda. This will be achieved by improving the radio-purity of close by components, as cables and holder materials, by increasing the detector mass by a factor 2-3 deploying novel inverted-coaxial detectors, by optimizing the HPGe and liquid argon array geometry reducing the number of neighbor strings and thereby improving the LAr light scintillation detection efficiency, and by improving the scintillation and optical properties of the liquid argon and hence the detectable amount of scintillation light. 

In the subsequent \Lthou~ stages  up to four additional pay loads will be deployed in a new liquid argon cryostat system. Each of the loads will hold between 200 and 300 kg of large mass inverted coaxial detectors with a total target mass of 1000~kg. The background index needs to be reduced by another factor of 10 with respect to \Ltwo~ to operate background-free for approximately 10~\tyr . The goal is to achieve a discovery sensitivity beyond $10^{28}$~years. In addition to the above mentioned measures, the \LEG ~ collaboration plans  to deploy the detector array in argon from underground production that is depleted in the isotope $^{42}$Ar to reach these background goals. Surface decays of the $^{42}$Ar progeny $^{42}$K could become a limiting background factor if not sufficiently mitigated by pulse shape analysis.  Also the alpha contamination on the passivated (small) surface area between p+ and n+ contacts shall be further reduced to allow sufficient margin to reach the background goals. 

The underground laboratory for \Lthou~ will be decided at a later stage. Under investigation are \Lngs , Boulby, SNOlab, SURF, CJPL and the new LSM, if realized timely. Preliminary design studies of the cryostat designs  have commenced considering the very space and access restrictions at the different host laboratories.  

\subsubsection{R\&D program and schedule}

The R\&D efforts for \Lthou~ build on the progress achieved for  \Ltwo . Specific \Lthou~ efforts focus on cryostat designs optimized for different underground sites, on the development of HPGe detectors with increased masses (from currently 2 to 3 kg) pertaining excellent pulse shape performance, on high-purity front-end electronics based on ASICs chips, on high-purity materials for cables and components close to the detectors, on liquid argon instrumentation with higher intrinsic purities, on the reduction of radioactive decays on the detector (n+) surface from  the $^{42}$Ar progeny $^{42}$K,  and on the reduction of alpha decays on the (p+ and grove) surfaces. Potentially limiting background events originating from  $^{42}$K will be reduced by the use of underground argon, currently under development by the INFN/DarkSide, and/or by improved signal recognition algorithms in conjunction with low-noise front-end electronics and optimized HPGe detector dead layer profiles.

Currently, there are two established industrial producers for $^{76}$Ge isotope enrichment and two established producers which can transform the germanium raw materials into state-of-the-art HPGe detectors. Additional detector producers are expected to be able to produce large mass HPGe diodes according to the \Lthou~ specifications in the near future.

The \LEG~ collaboration is currently preparing the upgrade of the \Gerda~ infrastructure for \Ltwo , producing new enriched detectors with masses around 2~kg, and preparing the hardware for commissioning the experiment in 2020/21. Assuming that the collaboration successfully secures the funding for \Lthou~by 2021, the design and construction can commence subsequently. Selection of the underground laboratory to host \Lthou~ will be carried out also at that time. Isotope procurement and detector production would start in parallel to the design and construction of the new experimental hardware infrastructures. Given the \Lthou~ cryostat system design, which can host up to four separate payloads, data taking can start with the first available detector payload allowing for an efficient staging of the experiment.

 The main costs are related to the isotope enrichment (40-50 M\euro /ton) and to detector production. Cryostat and infrastructure costs depend on the specific underground laboratory and can differ substantially.  For orientation, The \Gerda~ cryostat, water tank and main experimental infrastructures amounted to about 3~M\euro .


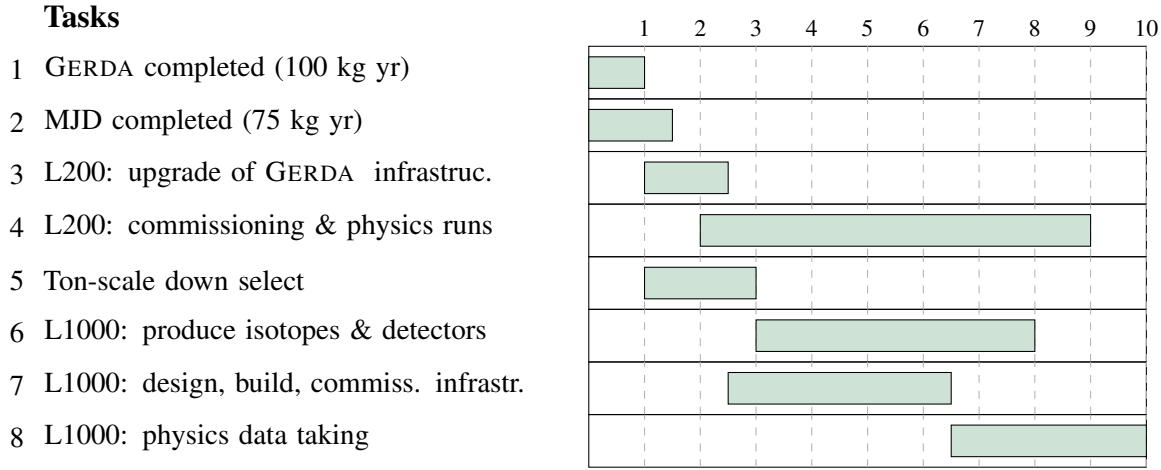
\begin{figure}[h]
    \begin{tikzpicture}
          \GanttHeader{0.9\textwidth}{2ex}{7.1cm}{10}
      \Task{1}{\Gerda~completed (100 kg yr) }{0}{1}
      \Task{2}{MJD completed (75 kg yr) }{0}{1.5}
    \Task{3}{L200:  upgrade of \Gerda~ infrastruc. }{1}{1.5}
     \Task{4}{L200:  commissioning \& physics runs }{2}{7}
      \Task{5}{Ton-scale down select }{1}{2}   
      \Task{6}{L1000:   produce isotopes \& detectors }{3}{5}
      \Task{7}{L1000:   design, build, commiss. infrastr.}{2.5}{4}
       \Task{8}{L1000:  physics data taking}{6.5}{3.5} 
    \end{tikzpicture}
    \caption{Schedule of the \LEG~ experiment with its main experimental infrastructure stages \Ltwo~ and \Lthou . Year 0 corresponds to January 2019 and the column numbers indicate years (ie. Year 2 = January 2021). Physics data taking of \Lthou~ could start earliest 2025/26 with the first detector payload of 200-300~kg of detectors, assuming a positive funding decision by 2021 (DOE down select). Isotope procurement and detector production continues in parallel for the deployment of the subsequent detector payloads. Payloads can be added with minimal interference with ongoing data taking.}
    \label{fig:gantt_Legend}
    \end{figure}
    

\subsection*{SWOT table: LEGEND ($^{76}$Ge)}
\begin{tcbraster}[raster columns=2, boxrule=0mm, arc=0mm]
\begin{tcolorbox}[equal height group=E, size=fbox, colback=swotS!60, colframe=swotS!80!black, title=\textsc{strengths}]
\begin{itemize}[label={$\bullet$}, topsep=0pt, itemsep=0pt, leftmargin=*]
\item HPGe diodes have best energy resolution (0.13\% FWHM) and lowest background achieved in ROI; prerequisite for signal discovery.
\item Background reduction of only a factor 6 for \Ltwo~ w.r.t. GERDA and factor 10 for \Lthou~  w.r.t. \Ltwo .
\item Efficient use of isotopes: total mass quasi equal to active mass given high signal acceptance efficiency.   
\item Efficient staging possible given design with separate payloads.
\item Wide availability of Ge; procurement has no impact on global market.
\item Two supplier for enrichment established and tested (Europe \& Russia).
\item  Comparative low spread of NME (factor ~2). 
\end{itemize}
\end{tcolorbox}
\begin{tcolorbox}[equal height group=E, size=fbox, colback=swotW!60, colframe=swotW!80!black, title=\textsc{weaknesses}]
\begin{itemize}[label={$\bullet$}, topsep=0pt, itemsep=0pt, leftmargin=*]
\item Requires deep underground laboratory and/or tagging for Ge-77m suppression. 
\item Underground Ar depleted in $^{42}$Ar likely required for \Lthou .
\item Relatively low Q-value (2039 keV) implies smaller phase space factor which requires larger $T_{1/2}$ for same values of $m_{\beta\beta}$.
\end{itemize}
\end{tcolorbox}
\begin{tcolorbox}[equal height group=F, size=fbox, colback=swotO!60, colframe=swotO!80!black, title=\textsc{opportunities}]
\begin{itemize}[label={$\bullet$}, topsep=0pt, itemsep=0pt, leftmargin=*]
\item LEGEND-200 start in 2021; \\ serves also as test bench for LEGEND-1000.
\item Non-DBD$0\nu$ physics at low energies.
\item Transatlantic cooperation and funding; \\ opportunities for new groups.

\end{itemize}
\end{tcolorbox}
\begin{tcolorbox}[equal height group=F, size=fbox, colback=swotT!60, colframe=swotT!80!black, title=\textsc{threats}]
\begin{itemize}[label={$\bullet$}, topsep=0pt, itemsep=0pt, leftmargin=*]
\item Unknown background could appear at \Ltwo~ which might be difficult to mitigate.
\item For \Lthou~: no funding secured; poor coordination of funding agencies; DOE down-select might move ahead without European funding aligned.
\item Underground argon production dependent on INFN/NSEF in context of DarkSide project.
\end{itemize}
\end{tcolorbox}
\end{tcbraster}

\subsection{Bolometers}
Bolometers are powerful low-energy particle detectors for the conduction of sensitive DBD0$\nu$-decay searches in the calorimetric approach~\cite{Fiorini:1984}. A bolometer consists of a single dielectric crystal --- the active part of the detector that contains the isotope of interest --- coupled to a temperature sensor. The signal, collected at very low temperatures ($< 20$~mK for large bolometers, with masses in the 0.1--1 kg range), consists of a thermal pulse registered by a dedicated sensor, with an amplitude of the order of 0.1~mK/MeV. 

The bolometric technique can provide high sensitive mass (via large detector arrays), high detection efficiency (70\%-90\%), high energy resolution (down to 0.15\%) and extremely low background thanks to potentially high material radiopurity and powerful methods to reject parasitic events~\cite{Poda:2017a}. Most of the favorable high $Q$-value DBD0$\nu$ decay candidates ($^{48}$Ca, $^{76}$Ge, $^{82}$Se, $^{96}$Zr, $^{100}$Mo, $^{116}$Cd, $^{124}$Sn, $^{130}$Te) can be studied with this technique. 

\subsubsection{Merits and limitations of CUORE}
An isotope of great interest for DBD0$\nu$ decay is $^{130}$Te. The signal is expected at 2527 keV, just below the end point of the $\gamma$ radioactivity at 2615~keV. The natural isotopic abundance of $^{130}$Te (34\%) is by far the highest among all the DBD0$\nu$-decay candidates. The experiment CUORE~\cite{Alduino:2018a} --- located in LNGS (Laboratori Nazionali del Gran Sasso) and currently in data taking --- consists of 988 TeO$_2$ bolometers (containing tellurium with a natural isotopic composition) with a mass of about 750~g each, corresponding to about 200~kg of $^{130}$Te. CUORE is one of the most sensitive DBD0$\nu$-decay experiments up-to-date. It has set a limit of $1.5 \times 10^{25}$~y on the half-life of $^{130}$Te, which leads to bounds of 75--350 meV on the effective Majorana mass~\cite{Alduino:2018a,Adams:2019}. The latter limit will be improved by about a factor 2 at the conclusion of the 4 years CUORE physics program. 

The background in the ROI of CUORE, corresponding to about 50 events/y, is dominated by energy-degraded $\alpha$ particles generated by surface contamination. They account for a background index $b$ of the order of \SI{E-2}{\ckky}~\cite{Alduino:2017a}. The CUORE background model~\cite{Ouellet:2018a}, built using directly the CUORE data, demonstrates that, after the elimination of the $\alpha$ component, $b =$~\SI{2.5E-3}{\ckky} is expected at $\sim 2.5$~MeV (in the ROI of $^{130}$Te), because of a $^{232}$Th contamination present in the cryostat inner thermal shields. On the contrary, $b =$~\SI{E-4}{\ckky} is safely estimated at energies higher than 2.6~MeV, where the contribution of $^{232}$Th is negligible. 

The CUORE cryostat is an unprecedented system in the field of cryogenics, which has represented a huge technological challenge. It took the CUORE collaboration quite some time --- around 2 years --- to understand and solve several technical issues. This pioneering work, now completed, has been very important to pave the way to large-scale bolometric experiments. A background run, with improved performance regarding the cryogenic system, is actually ongoing and also prove to run over some years. In conclusion, the CUORE cryostat has excellent performance and has demonstrated to be able to cool down one thousand of macro-bolometers to about 10 mK and successfully operate them~\cite{Buccheri:2014a}. 

\subsubsection{Scintillating bolometers}
Scintillating bolometers bring an additional value to the calorimetric technology. In these devices the crystal containing the isotope of interest is a scintillator, and a second auxiliary bolometer to register the emitted scintillation light is operated close to it. The ensemble of the crystal containing the DBD0$\nu$-candidate and its light detector is referred to as detector module. The simultaneous detection of heat and scintillation light allows one to distinguish $\alpha$ particles from electrons or $\gamma$'s thanks to the different light yield and signal shape~\cite{Pirro:2006a}, eliminating the dominant background source observed in CUORE. Scintillating bolometers containing a candidate with a Q-value higher than 2615 keV have therefore the potential to provide a background-free search even at a ton$\times$year exposure~\cite{Artusa:2014a}. Candidates that fit the high-Q-value requirement and can be embedded in scintillating crystals are $^{82}$Se (Q-value~=~2998~keV, compound ZnSe), $^{100}$Mo (Q-value~=~3034~keV, compound Li$_2$MoO$_4$) and $^{116}$Cd (Q-value~=~2813~keV, compound CdWO$_4$). 

In spite of the excellent results achieved by the demonstrator CUPID-0~\cite{Azzolini:2018a} in LNGS (which has established the most stringent current limit on the DBD0$\nu$-half-life of $^{82}$Se by using enriched Zn$^{82}$Se crystals), some drawbacks also appeared for ZnSe. This compound has a difficult crystallization procedure, features a good but not excellent energy resolution (10--30 keV FWHM) and an internal contamination in $^{228}$Th at the level of some tens of $\mu$Bq/kg at the present state of the technology. Good results were achieved also on $^{116}$Cd with enriched $^{116}$CdWO$_4$ crystals at the prototype level~\cite{Barabash:2016a}, but the isotope $^{116}$Cd is intrinsically more difficult to enrich with respect to the other candidates, with a consequent higher cost by a factor $\sim 2$.

For all these reasons, an intense R\&D activity has focused on $^{100}$Mo-containing compounds, and in particular on Li$_2$MoO$_4$~\cite{Bekker:2016a,Armengaud:2017a}, mainly in the framework of the LUMINEU project. The achieved results have been confirmed on an intermediate scale by the currently running CUPID-Mo demonstrator (consisting of 20 modules of 210~g each)~\cite{Poda:2017b,Poda:2018a}, installed in the Modane underground laboratory (LSM). The single module of CUPID-Mo consists of a crystal of Li$_2^{\ 100}$MoO$_4$ enriched at more than 95\% in $^{100}$Mo. The crystal is a cylinder with 44~mm diameter and 45~mm height coupled to an NTD (Neutron Transmutation Doped) Ge thermistor. At least one of the flat surfaces is exposed to a light detector, consisting of a NTD-instrumented Ge wafer ($\oslash$=44~mm, thickness=0.17 mm) and coated with a 70-nm-thick SiO layer on both sides to maximize light absorption. The results achieved demonstrate the maturity reached by the proposed technology and the high standard of the detectors~\cite{Armengaud:2017a,Poda:2017b,Poda:2018a}: energy resolutions of $\sim$5~keV FWHM at 2615 keV have been routinely obtained; $\alpha$ rejection at the level of 99.9\% has been obtained thanks to the heat-light readout; minimal internal contamination (inferior to $\sim$~5~$\mu$Bq/kg for both $^{232}$Th and $^{238}$U as well as to 5~mBq/kg for $^{40}$K) were guaranteed by detailed crystal production protocols.

These promising achievements, along with the results of the CUORE background model~\cite{Alduino:2017a,Ouellet:2018a}, have lead the CUORE, CUPID-0 and CUPID-Mo collaborations to the decision to select the Li$_2$MoO$_4$ technology for the future CUPID experiment.

\subsubsection{CUPID}
CUPID (CUORE Upgrade with Particle IDentification) is a proposed next-generation DBD0$\nu$-decay experiment based on scintillating bolometers to be installed in the cryogenic infrastructure currently hosting CUORE at LNGS~\cite{CUPID:2019a}. The bolometer crystals will be grown from Li$_2^{\, 100}$MoO$_4$ enriched to 95\% in $^{100}$Mo. At the present stage of the conceptual design, the CUPID collaboration envisions cylindrical crystals with 50~mm diameter and 50~mm height, corresponding to a mass of 301 g each. The flat surfaces of the crystals will be exposed to bolometric light detectors fabricated from Ge wafers with 5 cm diameter, using an NTD Ge thermistor as a thermal sensor. The crystals will be stacked in detector towers conceptually similar to those of CUPID-0~\cite{Azzolini:2018a} and CUPID-Mo~\cite{Poda:2018a}. With this design, about 1500 crystals will be hosted by the CUORE cryostat, corresponding to about 250 kg of $^{100}$Mo. Note that the size of the main crystal and of the light detector, their mechanical and geometrical arrangements, and the readout approach proposed for CUPID closely resemble the configurations successfully adopted in CUPID-0 and CUPID-Mo. 

The main strengths of CUPID are the following:
\begin{itemize}[noitemsep,topsep=0pt]
\item The single-module technology and the related production and purification protocols are fully established~\cite{Armengaud:2017a,Berge:2014a,Grigorieva:2017a,Poda:2018a}.
\item The infrastructure for CUPID exists and is operational~\cite{Alduino:2019a}, even though it needs some upgrades discussed below.
\item Most of the detector assembly protocols adopted in CUORE can be extended to CUPID with minor modifications~\cite{Buccheri:2014a}.
\item The combination of the results achieved in LUMINEU and CUPID-Mo with the CUORE background model allows to predict a background index of the order of \SI{E-4}{\ckky} in the ROI of $^{100}$Mo~\cite{CUPID:2019a}.
\end{itemize}

Some R\&D is still required in order to achieve safely $b =$~\SI{E-4}{\ckky}: the CUPID background goal. In particular, pile-up rejection methods need to be refined in order to control the contribution to the background coming from random coincidences of $^{100}$Mo ordinary $2\nu\beta\beta$ decay events~\cite{Chernyak:2012a,Chernyak:2014a}. An improvement of the signal-to-noise-ratio in light detectors with respect to the state of the art could be required~\cite{Chernyak:2016a}. Secondly, residual surface radioactivity in the reflective foil surrounding the scintillating crystals can be challenging with respect to the CUPID background target. The most straightforward solution is to avoid completely the foil and minimize the amount of inert material between crystals, but this will reduce the light collection. Dedicated tests will fix the optimal configuration.

The CUORE cryostat will need upgrades. The wiring must be extended in order to read out $\sim 3000$ channels instead of the current $\sim 1000$. This operation looks straight-forward and no major problem is envisaged. Light detectors are particularly sensitive to vibrations and must operate in a much lower energy range with respect to the main DBD0$\nu$-decay crystals. This may require additional studies and interventions to improve the isolation of detectors from vibrations. 

The 3$\sigma$ CUPID discovery sensitivity on $m_{\beta\beta}$ is 12--20~meV in 10~y live time.

Upgrades are possible beyond the currently proposed version of CUPID. The present background model indicates the directions to be taken in order to reduce the background index by at least further order of magnitude, bringing it to the level of $b \sim$~\SI{E-5}{\ckky} or less. R\&D activities are planned in this prospect.  The CUPID collaboration is therefore discussing two possible scenarios after CUPID. The former is exactly the detector considered in CUPID, but operating in a nearly zero-background mode, which corresponds to the background index of $2\times10^{-5}$\ckky. The latter is an ultimate bolometric detector, CUPID-1T, consisting of 1.8~tons of Li$_2$MoO$_4$, or 1000~kg of $^{100}$Mo. Such detector could be accommodated in a new cryostat approximately 4 times larger than CUORE. For optimal sensitivity, the background should be further reduced to the level of $5\times10^{-6}$\ckky. This is very challenging, but within the realm of
possibility for the transition energy of 3034~keV. The 3$\sigma$ half-life discovery sensitivity of these two future searches would be $2\times10^{27}$~y and $8\times10^{27}$~y respectively in 10~y live time.

\subsubsection{CUPID schedule}
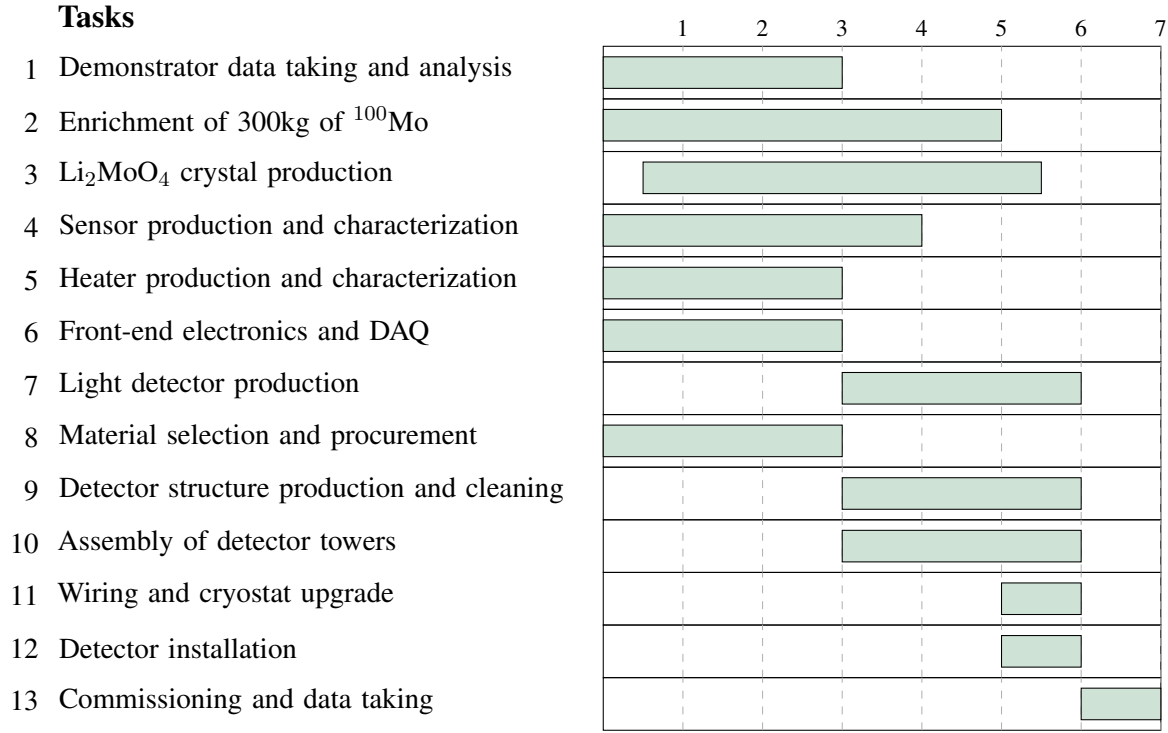
\begin{figure}[h]
    \begin{tikzpicture}
      \GanttHeader{0.9\textwidth}{2ex}{7.1cm}{7}
      \Task{1}{Demonstrator data taking and analysis}{0}{3}
      \Task{2}{Enrichment of 300kg of $^{100}$Mo}{0}{5}
      \Task{3}{Li$_2$MoO$_4$ crystal production}{0.5}{5}
      \Task{4}{Sensor production and characterization}{0}{4}
      \Task{5}{Heater production and characterization}{0}{3}
      \Task{6}{Front-end electronics and DAQ}{0}{3}
      \Task{7}{Light detector production}{3}{3}
      \Task{8}{Material selection and procurement}{0}{3}
      \Task{9}{Detector structure production and cleaning}{3}{3}
      \Task{10}{Assembly of detector towers}{3}{3}
      \Task{11}{Wiring and cryostat upgrade}{5}{1}
      \Task{12}{Detector installation}{5}{1}
      \Task{13}{Commissioning and data taking}{6}{1}
    \end{tikzpicture}
    \caption{Time schedule of the CUPID experiment. Column numbers indicate years.}
    \label{fig:gantt_Cupid}
    \end{figure}

The CUPID program is reported in Fig. \ref{fig:gantt_Cupid}, in the assumption of fully and promptly available funding.

\subsubsection{CUPID collaboration and resources}
The CUPID collaboration builds on the CUORE, CUPID-0 and CUPID-Mo collaborations but is open to new participants. It comprises currently about 130 members from six countries. It is expected that the funding will come mainly from three countries: Italy, France and US. A strong effort is ongoing in China to develop locally a scintillating-bolometer technology, with a possible CUPID-China experiment in the CJPL underground laboratory, complementary to CUPID at LNGS.

The CUPID pre-CDR document was submitted to arXiv~\cite{CUPID:2019a}  in July 2019 and it was also presented and evaluated by the scientific committee of the Gran Sasso Laboratory during the meeting of October 2019. The CUPID Technical Design Report is foreseen at the end of 2020.

Since the infrastructure already exists, the cost is dominated by the enrichment (15--20 M\euro), the crystallization (4 M\euro), the detector assembly and the cleaning (3 M\euro) and the upgrade of the electronics and DAQ (4 M\euro). Some possible other costs will be related with some small upgrades and with the general maintenance of the cryogenic infrastructure. The activities to define the detector structures (2019--2021) are funded in Italy, France and US. In the latter country, CUPID has been selected by DoE at the CD0 level. Funding for enrichment and crystallization are not secured yet.

\subsection*{SWOT table: CUPID ($^{100}$Mo)}
\begin{tcbraster}[raster columns=2, boxrule=0mm, arc=0mm]
\begin{tcolorbox}[equal height group=C, size=fbox, colback=swotS!60, colframe=swotS!80!black, title=\textsc{strengths}]
\begin{itemize}[label={$\bullet$}, topsep=0pt, itemsep=0pt, leftmargin=*]
\item Enrichment at large scale with medium prices
\item High Q-value (3034 keV)
\item Compatible with scintillating bolometer technique
\item Excellent energy resolution\\
Li$_2^{\ 100}$MoO$_4$: 5 keV FWHM at 2615 keV
\item Low background demonstrated in large crystal: 
$\sim$~5~$\mu$Bq/kg for $^{232}$Th / $^{238}$U;
5~mBq/kg for $^{40}$K
\item Source=Detector, modularity, high efficiency
\item Event-type discrimination: \\   $\alpha/\beta$ full rejection demonstrated
\item Favourable Nuclear Factor of Merit \\
(Phase Space x NME)
\end{itemize}
\end{tcolorbox}
\begin{tcolorbox}[equal height group=C, size=fbox, colback=swotW!60, colframe=swotW!80!black, title=\textsc{weaknesses}]
\begin{itemize}[label={$\bullet$}, topsep=0pt, itemsep=0pt, leftmargin=*]
\item No tracking
\item Short 2$\nu$2$\beta$ half-life \\ (potential background due to accidental pileup) \\ $\Rightarrow$ develop faster light detector
\item Scalability possible but costly; \\ factor two looks feasible by setting up a second CUORE-like facility
\item Cryogenic infrastructures are complicated and need onsite expertise
\end{itemize}
\end{tcolorbox}
\begin{tcolorbox}[equal height group=D, size=fbox, colback=swotO!60, colframe=swotO!80!black, title=\textsc{opportunities}]
\begin{itemize}[label={$\bullet$}, topsep=0pt, itemsep=0pt, leftmargin=*]
\item Cryogenic infrastructure well demonstrated in CUORE (space for 300 kg of $^{100}$Mo-enriched detector available) 
\item Several crystal compounds compatible with the bolometric technique: \\ Li$_2^{\ 100}$MoO$_4$, ZnMoO$_4$, CaMoO$_4$
\item High reproducibility of crystal quality 
\item Many producers on the market
\item Alternative pulse shape discrimination techniques
\item Second physics case (direct dark matter detection)
\item New CUPID collaboration is chance for new \\ collaborators/groups
\end{itemize}
\end{tcolorbox}
\begin{tcolorbox}[equal height group=D, size=fbox, colback=swotT!60, colframe=swotT!80!black, title=\textsc{threats}]
\begin{itemize}[label={$\bullet$}, topsep=0pt, itemsep=0pt, leftmargin=*]
\item Enrichment monopoly in Russia
\item AMORE collaboration: \\  120 kg of $^{100}$Mo for bolometric experiment \\ in Korea. \\  This can be turned into an opportunity in case of \\ a common CUPID-AMoRE bi-site experiment
\item Funding of CUPID open
\end{itemize}
\end{tcolorbox}
\end{tcbraster}

\subsection{Xenon TPC}

\subsubsection{State of the art}
Over the last decade, Xenon TPCs (XeTPC) have emerged as powerful tools for the study of rare events, in particular concerning dark matter and DBD0$\nu$ searches. 
In a XeTPC, charged radiation ionizes the fluid and the ionization electrons are drifted under the action of an electric field to sensitive image planes, where their transverse position information X,Y is collected. Their arrival times (relative to the start-of-the-event time, or \tz) are then traded to longitudinal positions, Z, through their average drift velocity. 

In the case of DBD0$\nu$ searches, Xenon is not only the sensitive medium, but also the target where the decays occur. Since the sensitivity of the search is proportional to the target mass the apparatus needs to be as large and compact as possible, leading to either high pressure Xenon (HPXe) or liquid Xenon (LXe) TPCs.  Both types of detectors act as calorimeters, capable of measuring the total energy of the decay and to locate the interaction in a well defined fiducial volume thanks to the availability of a mechanism to signal \tz, namely the VUV scintillation emitted by Xenon as a response to ionizing radiation. 
In addition a HPXe TPC provides a {\em topological signature}, thanks to its capability to image the electron tracks.


The suitability of LXe TPCs for DBD0$\nu$ decays has been demonstrated by the EXO experiment, which has reached a sensitivity to \Tonu\ of \SI{3.7E+25}{\yr} ~\cite{Albert:2017owj}. The proposed next-stage for EXO would be the nEXO apparatus, a 5-ton liquid xenon TPC whose projected sensitivity to \Tonu\ would reach \SI{9.2E+27}{\yr} in a 10 year run \cite{Albert:2017hjq}. 
A characteristic of LXe TPCs is that their performance for DBD0$\nu$ searches improves a priory as the detector becomes larger as self-shielding improves. This is both an asset (since it makes feasible to build monolithic, very massive detectors) and a liability, since it is difficult to stage the apparatus building smaller, cheaper modules. 


Although not as massive, HPXe TPCs offer two distinctive advantages over LXe: better energy resolution (if electroluminescence is used to amplify the signal) and the capability to separate the two electrons emitted in a DBD0$\nu$ decay from single electron, and multi-energy deposition backgrounds. In addition, recent developments show that the in-situ identification of the barium ion produced in the xenon DBD0$\nu$ decay (in delayed coincidence with the measurement of the two-electron signature) may be feasible in a relatively short time. In Europe, the technology of electroluminescent high pressure xenon gas TPCs (\HPXeEL) for DBD0$\nu$ searches is being developed by the 
NEXT  program ~\cite{Nygren:2009zz, Gomez-Cadenas:2013lta, Martin-Albo:2015rhw}. 

\subsubsection{\Next}
 
The \Next\  detector, scheduled to start data taking in 2020, is a radiopure asymmetric \HPXeEL\ deploying 100~kg of xenon. The fiducial region is a cylinder of \NextTpcDiameter\ diameter and \NextTpcLength\ length, (\NextFiducialVolume\ fiducial volume) holding a mass of \NextFiducialMass\ xenon gas enriched at \XeEnrichment\ in \XE, and operating at \NextPressure.  
The electroluminescent light is detected by two independent sensor planes. 
The energy of the event is measured by integrating the amplified EL signal (\stwo) with a {\em energy plane}  (EP) featuring \NextNumberOfPMT\ photomultipliers (PMTs). In addition the EP records the \sone\ signal which provides the \tz\ of the event.  
EL light is also detected a few mm away from production at the anode plane by a dense array of silicon photomultipliers (featuring \NextNumberOfSiPM\ SiPMs) known as the  \emph{tracking plane}. This measurement allows for topological reconstruction since it provides position information transverse to the drift direction. 

\Next\ is the third phase of a program which started in 2009 with the construction of the NEXT-DBDM and NEXT-DEMO  prototypes, which demonstrated the robustness of the technology, its excellent energy resolution and its unique topological signal \cite{Alvarez:2012xda, Alvarez:2013gxa, Alvarez:2012hh, Ferrario:2015kta}. The \NEW\ demonstrator~\cite{Monrabal:2018xlr}, a scale model of \Next\ implements the second phase of the program. \NEW\ is a radiopure detector, deploying 5 kg of xenon, which has been taking data at the underground laboratory of Canfranc (LSC) since 2016. Operation of \NEW\  established a procedure to calibrate the detector with krypton decays\cite{Martinez-Lema:2018ibw}, and provided initial measurements of energy resolution~\citep{Renner:2018ttw}, electron drift parameters~\cite{Simon:2018vep} and a measurement of the impact of \RAD\ in the radioactive budget \cite{Novella:2018ewv}. 

Recent results of \NEW\ include the measurement of an energy resolution at \Qbb\ better than 1\% FWHM~\cite{Renner:2019pfe}, the demonstration from the data themselves of a robust discrimination between 2-electrons (which characterize a double beta decay) and single background electrons~\cite{Ferrario:2019kwg}, and a measurement of the backgrounds, which demonstrates both the low radioactive budget of the apparatus and the good quality of the background model \cite{NewRadiogenic}. 

The combination of good energy resolution, topological discrimination  and low radioactive budget, results in a very low expected background index of \SI{4E-4}{\ckky} \cite{Martin-Albo:2015rhw}. This results into a projected background in the ROI for \Next\ of $<$0.7~counts~yr\ensuremath{^{-1}}, which translated into a sensitivity of \SI{1E+26}{\yr} after a total exposure of 400 kg$\cdot$y.  
\subsubsection{NEXT-HD}
The \HPXeEL\ technology can be scaled up to multi-ton target masses introducing several technological advancements already available \citep{JJMoriond:2019}. The most important innovation is the replacement of PMTs (which are the leading source of background in \Next) with SiPMs, which are intrinsically radiopure, resistant to pressure and able to provide better light collection. Furthermore it is possible to  optimize the topological signature performance through the operation of the detector with low diffusion mixtures \cite{Henriques:2017rlj, Felkai:2017oeq, McDonald:2019fhy, Renner:2017ey}, resulting in better position resolution and, thus, improving the signal-background separation \cite{Renner:2017ey}. 
The incremental approach to ton-scale \HPXeEL\ detectors is called ``high definition" (HD). Monte Carlo simulations show that the specific background rate of \Next\ may be reduced by at least one order of magnitude, reaching a background index of $5 \times 10^{5}$ \ckky. 
A NEXT-HD module with a mass in the ton range will be able to improve by more than one order of magnitude the current limits in \Tonu, thus exceeding $\Tonu > 10^{27}$~yr. 
 
\subsubsection{NEXT-BOLD}
The \HPXeEL\ technology may permit even a more radical approach to the next generation of DBD0$\nu$ experiment by implementing a system capable of detecting with high efficiency the presence of the \Bapp\ ion produced in the \XE\ DBD0$\nu$ decay. The detection would occur in (delayed) coincidence with the identification of the two electrons and would ensure a background free experiment. 

Single Molecular Fluorescence Imaging (SMFI) uses molecules which include a fluorescent group, called fluorophore and  a metal binding group which inhibits fluorescence unless the molecules is chelated with a suitable ion. 
 The possibility of using SMFI as the basis of molecular sensors for barium tagging was proposed in \cite{nygrenbata, Jones:2016qiq}, followed,  shortly after, by a proof of concept which managed to resolve individual 
\Bapp\ ions on a scanning surface using an SMFI-based sensor \cite{McDonald:2017izm}. 

Intense R\&D is under way in the NEXT collaboration to implement a SMFI sensor prototype capable to demonstrate barium detection. New molecular indicators, able to provide a intense fluorescence signal in dry medium were presented recently \cite{Thapa:2019zjk}. Even more recently, bicolor fluorescent indicators (FBIs) have been developed \cite{Rivilla:2019vzd}. FBIs add an extra handle to conventional indicators, shifting the spectrum of chelated indicators with respect to the spectrum on unchelated molecules. The result is a very large separation factor, in excess of $10^4$ between chelated and unchelated species, which makes the prospect of a barium target much more realistic. Although many steps need to be taken to demonstrate a full barium tagging sensor, the consistent success of the R\&D initiated in 2016 yields good prospects. If the efforts under way succeed, the possibility to build a ton-scale, background-free experiment based in the \HPXeEL\ technology becomes very appealing. This disruptive approach is called  ``Barium On Light Detection'' (BOLD). A NEXT-BOLD module  would measure the energy and event position in the anode (with a SiPM plane) reserving the cathode for the barium sensor. The delayed coincidence would permit relaxing the stringent topological restrictions imposed to the events in NEXT-100 (and NEXT-HD), resulting in a higher signal efficiency in addition to a negligible background index. A NEXT-BOLD module with a mass in the ton range could reach a sensitivity $\Tonu > 10^{28}$~yr. 

\subsubsection{R\&D program and schedule}

\begin{figure}[h]
    \begin{tikzpicture}
      \GanttHeader{0.9\textwidth}{0.2}{7.5cm}{10}
      \Task{1}{Operation of NEXT-White}{0}{1}
      \Task{2}{Assembly and commissioning of NEXT-100}{0}{1}
      \Task{3}{Operation of NEXT-100}{1}{3}
      \Task{4}{R\&D for NEXT-HD and NEXT-BOLD}{1}{3}
       \Task{5}{Choice of technology for ton-scale module}{3}{1}
      \Task{6}{Construction of first ton-scale module}{4}{1}
      \Task{7}{Procurement of isotope mass}{2}{3}
      \Task{8}{Operation of first ton-scale module}{5}{3}
      \Task{9}{R\&D for second ton-scale module}{5}{3}
      \Task{10}{Choice of technology}{7}{1}
      \Task{11}{Construction of second ton-scale module}{8}{1}
      \Task{12}{Procurement of isotope mass}{6}{3}
      \Task{13}{Operation of two ton-scale modules}{9}{1}
    \end{tikzpicture}
    \caption{Time schedule of NEXT experiment including R\&D program. Column numbers indicate years. Year 0 is 2019. Operation of two modules continue after year 10.}
    \label{fig:gantt2}
    \end{figure}
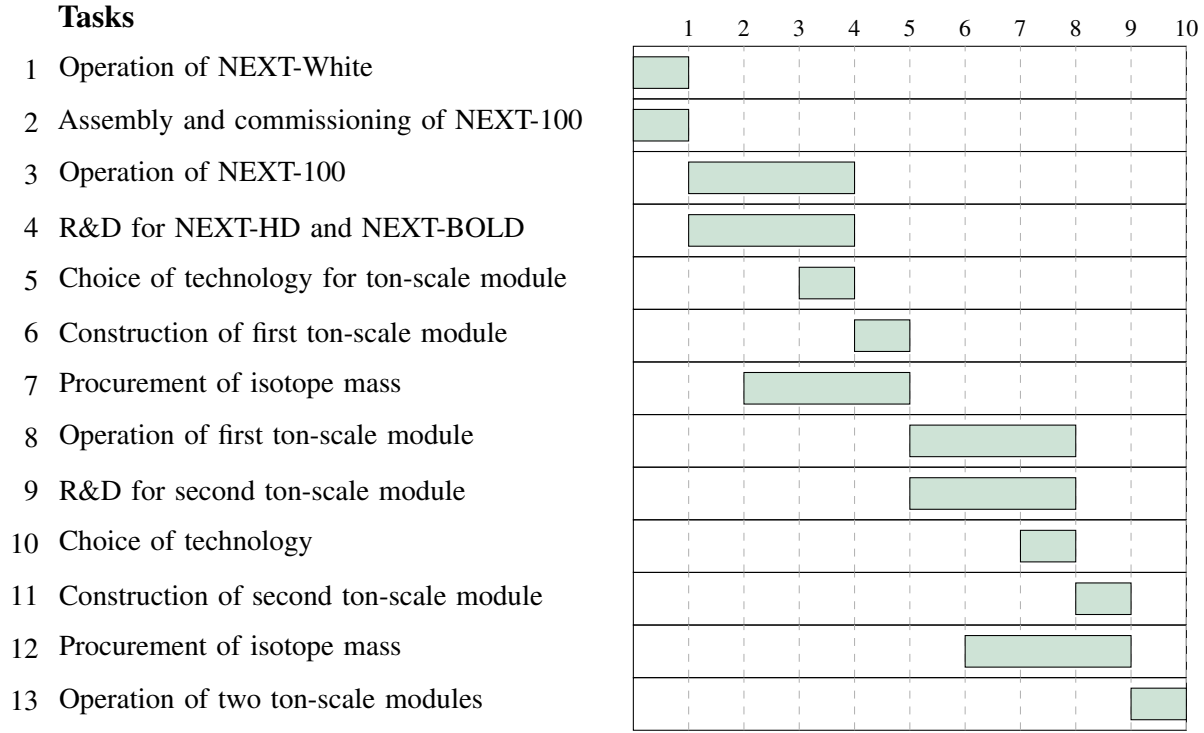

The NEXT program contemplates the construction of two ton-scale modules. At present we assume that each module will have a mass of 1 ton, although a likely scenario is that the first module will have a lighter mass (in the range of 500--750 kg) and the second module a larger mass (1250--1500 kg). The R\&D between 2020 and 2023 will decide the characteristics of the first module (dimensions, pressure, temperature), will address the operational challenges (gas purity, electric fields), design the construction of the sensor planes and choose the technology (HD or BOLD). The R\&D between 2025 and 2028 will focus on increasing the performance of the second module. If the chosen technology is HD, the focus of the R\&D will be to reduce the radioactive budget. If BOLD can be implemented, the focus will instead be to increase the selection efficiency. If \Bapp\ tagging can be turned into a real possibility during the next few years, the baseline scenario assumes that the first module will be NEXT-HD and the second will be NEXT-BOLD. 

\subsubsection{Collaboration and resources}
 At present the NEXT collaboration includes about 80 scientists working in 21 universities and laboratories from Spain, United Stated, Portugal, Israel, Colombia and Russia. The funding agencies of the international collaboration, as well as the ERC (though an AdG/ERC granted to J.J. Gomez-Cadenas) contribute to the funding of the program. The main contribution comes from Spain, followed by USA. At present the \Next\ detector construction and operation is fully funded. There is also significant funds for R\&D during the next few years. Furthermore, we intend to explore the potential synergy with the Dark Side detector at LNGS, which intends to build a SiPM-only argon TPC, similar in many aspects to the proposed ton-scale NEXT modules. 
 
 Funding for the construction of the first ton-scale module will be sought out from the existing funding agencies, the ERC and potential new members. Funding procurement for the ton-scale detector will start at earliest after a full demonstration of the feasibility of the technology, that is, after the construction and initial operation of NEXT-100. 
 
 The cost of the NEXT-100 detector can be divided in three major items: enriched gas, apparatus and infrastructure (shielding and gas system). The approximate cost of each item has been 1 M\euro. For each ton-scale module, we expect the cost of the gas to scale linearly (thus 10M\euro per ton of isotope). The cost of the detector will be dominated by the SiPMs, a technology which keeps reducing costs yearly. We estimate 2.5M\euro for each ton-scale detector and 2.5M\euro for the infrastructures. Each ton-scale module would then cost about 15M\euro.

 \subsection*{SWOT table: NEXT ($^{136}$Xe)}
\begin{tcbraster}[raster columns=2,boxrule=0mm, arc=0mm]
\begin{tcolorbox}[equal height group=A, size=fbox, colback=swotS!60, colframe=swotS!80!black, title=\textsc{strengths}]
\begin{itemize}[label={$\bullet$}, topsep=0pt, itemsep=0pt, leftmargin=*]
\item Enrichment at large scale with low prices \\ (10 M\euro\ per ton)
\item Moderately high Q-value (2457 keV)
\item Long 2$\nu$2$\beta$ half-life
\item Good energy resolution \\
NEXT-White: 20 keV FWHM at 2457 keV
\item NEXT-White: factor 20 reduction in background due to topological cuts. 
\item Source=Detector.
\item Fiducial volume: only high energy gammas relevant, negligible background from $\alpha$ 
\item Reasonable Nuclear Factor of Merit \\
(Phase Space x NME)
\item Possibility of in-situ barium tagging, leading to a background-free experiment
\end{itemize}
\end{tcolorbox}
\begin{tcolorbox}[equal height group=A, size=fbox, colback=swotW!60, colframe=swotW!80!black, title=\textsc{weaknesses}]
\begin{itemize}[label={$\bullet$}, topsep=0pt, itemsep=0pt, leftmargin=*]
\item Modest/low efficiency (30\%) 
\item Less dense than liquid xenon
\item Maximum size of modules about 500-1500 kg \\  $\Rightarrow$ Possibility to build two modules
\item Less developed than other DBD0$\nu$ technologies
\item Physics potential (background index, barium tagging) still under investigation.
\end{itemize}
\end{tcolorbox}
\begin{tcolorbox}[equal height group=B, size=fbox, colback=swotO!60, colframe=swotO!80!black, title=\textsc{opportunities}]
\begin{itemize}[label={$\bullet$}, topsep=0pt, itemsep=0pt, leftmargin=*]
\item Full infrastructure for operation of NEXT-100 and possible upgrades available at  Canfranc Underground Laboratory
\item NEXT-100 is a high profile scientific project in Spain
\item US plays an important role in NEXT. 
\item Possibility of a major future US participation.
\item Possibility of reusing other major infrastructures (BOREXINO at LNGS) for ton-scale modules
\item Interest in HPXe in Japan (and China) with the possibility of convergence  
\item Potential synergy with dark matter experiments (Dark Side, DARWIN)

\end{itemize}
\end{tcolorbox}
\begin{tcolorbox}[equal height group=B, size=fbox, colback=swotT!60, colframe=swotT!80!black, title=\textsc{threats}]
\begin{itemize}[label={$\bullet$}, topsep=0pt, itemsep=0pt, leftmargin=*]
\item Xenon market potentially overloaded (dark matter experiments, neXO)
\item Funding not yet guaranteed beyond NEXT-100
\item Intense competition with other projects may lead to the technology not being selected in the US
\item Interest in HPXe in Japan (and China) but possibility of {\em no} convergence.
\end{itemize}
\end{tcolorbox}
\end{tcbraster}

 \subsubsection{Possible future synergy with direct Dark Matter searches }
 The next-generation HPXe detectors, (HD or BOLD) will not have PMTs. In particular, NEXT-HD envisions a cool gas detector instrumented only with VUV-sensitive SiPMs , that will be used both for energy measurement and track reconstruction. 
 This is exactly the same scheme that is being implemented by the Dark Side collaboration for the construction of the 
 Dark Side 20 ton detector. The synergy is obvious, with the possibility of sharing the NOA facility currently being commissioned at L'Aquila. 
 
 The intriguing possibility that DARWIN may have a sensitivity to DBD0$\nu$ decays similar to that of nEXO, also opens an exciting scenario for European lead DBD0$\nu$ searches. Notice, first, the complementary of the techniques: NEXT is primarily a DBD0$\nu$ search technology which may have an impact in Dark Matter searches while DARWIN is a dark matter experiment which may impact DBD0$\nu$ searches. Furthermore, the experiments investigate the same isotope with different approaches (topology and excellent energy resolution, with potential barium-tagging in the case of NEXT, self-shielding and good energy resolution with large mass in the case of Darwin) and thus, any potential signal could be thoroughly cross-checked. Last but not least, wide collaboration, in particular at the level of material screening and selection (which will be critical for both detectors) can be established. 
 
 \subsection{International context}
 
 \subsubsection{AMoRE}
The AMoRE (Advanced Mo-based Rare process Experiment) experiment aims to search for DBD0$\nu$ decay of $^{100}$Mo using molybdate scintillating crystals operating at cryogenic temperatures. The complete AMoRe setup will run 200 kg of XMoO4 crystals at about 10 mK temperature: X is a placeholder for $^{40}$Ca, Li, or Na atoms. For the final detector configuration, AMoRE experiment aims for an energy resolution better than 10 keV FWHM, and a total background rate in the region-of-interest (around 3043 keV) lower than 10$^{-4}$ counts/(keV·kg·year).

A pilot stage of the experiment using six $^{40}$Ca$^{100}$MoO$_{4}$ crystals (total mass 1.9 kg) was performed at Yangyang underground laboratory (South Korea). In AMoRE Metallic Magnetic Calorimeters (MMCs) are employed as phonon sensors to measures the temperature rise of the crystal induced by the radiation absorption. An auxiliary light absorber, also equipped with a MMC sensor, detects the amount of scintillation light produced in the crystal. The achieved FWHM energy resolutions at the Q-value of the $^{100}$Mo $\beta \beta$ decay are between 10--17 keV, the measured background level is 0.55 counts/(keV kg year). 

To improve the detector performance AMoRE collaboration is working on different types of $^{100}$Mo-enriched crystals, in particular Li$_{2}$MoO$_{4}$ and Na$_{2}$MoO$_{4}$. The internal background level and achievable detector performance will be investigated to finally select the type of crystals to be used for the full detector configuration. 
AMoRE aims at improving the effective Majorana neutrino mass sensitivity to 20-50 meV. The experiment is planned to be installed in Yemi lab, a new 1,000 m deep underground site presently under construction in South Korea and that will be available in 2020.
 
 \subsubsection{HPXE}
 In addition to the NEXT program, the HPXe technology is being pursued in China and in Japan. 
 
 In China, the PANDAX-III collaboration \citep{Chen:2016qcd, Han:2017fol}
 proposes an experiment at the China Jin-Ping underground Laboratory II (CJPL-II). The first phase of the experiment would be a TPC with a target mass of 200 kg of xenon enriched at 90\% in \XE\ and operating at 10 bar. Unlike NEXT, PANDAX-III pursues a readout based on Microbulk Micromegas, a fine pitch micro-pattern gas detector. The detector will operate with a Xe-TMA mixture which, at the same time, reduces the natural diffusion of xenon and suppresses the primary scintillation, resulting therefore in a detector without start-of-the-event ($t_0$) signal. As found by the St. Gotthard experiment \citep{Wong:1991vd, Luscher:1998sd} and recently confirmed by the analysis of radiogenic backgrounds in \NEW\citep{NewRadiogenic} this represents a serious handicap, since suppressing backgrounds coming from the end-caps of the detector (in particular from the cathode) becomes much more difficult in the absence of the fiducialization provided by $t_0$. The energy resolution claimed by the technology, is 3\% FWHM. The claim is based in studies of the NEXT collaboration using a 1-kg detector \citep{Gonzalez-Diaz:2015oba}. In those studies, the energy resolution of 1.275 MeV long electron tracks was measure with an energy resolution of 4.6 \%, which extrapolates (assuming $1/\sqrt{E}$~scaling, to 3.3\% FWHM. The energy measurement has not been made at \Qbb\ and the scaling assumptions, therefore, need yet to be confirmed by data. 
 
 In exchange of the absence of $t_0$ and the poorer energy resolution, the use of low diffusion mixtures combined with a fined-grained readout offers an excellent performance of the topological signature, as shown by Monte Carlo studies. For example, the performance obtained in \citep{Qiao:2018edn}, using convolutional neural networks (CNNs) is of the order of 90\% for the signal efficiency and about 95\% background rejection, quite close to the results found by the NEXT collaboration in a previous Monte Carlo study using CNNs \citep{Renner:2017ey}. Indeed, part of the improvement expected in the NEXT-HD detector comes from an improved topological signature. Notice, however, that, unlike the case of NEXT \citep{Ferrario:2015kta, Ferrario:2019kwg}, the performance of the PandasX-III detector has not yet been validated with data. 
 
 In Japan, the AXEL R\&D 
 lead by the U. of Kyoto, is developing new concepts for electroluminescent HPXe TPCs \citep{Ban:2017ett}. The envisioned AXEL detector is almost identical to the NEXT design. Both are HPXe-EL TPCs, with an energy plane (AXEL assumes PMTs, as the \Next\ detector) and a tracking plane based in SiPMs. In the AXEL concept, however, the SiPMs are VUV sensitive and provide a measurement of the energy of the event in addition of a reconstruction of the topological signal. Initial results are available from small prototypes, showing an energy resolution which is still not competitive with that achieved by NEXT. 
On the other hand, the AXEL R\&D addresses two important points. One is the need to build very large EL structures for future ton-scale detectors. In that respect, the modular nature of their proposed tracking plane appears  
well suited to scale up to large dimensions. The second point is 
the interest to measure the energy at the anode, if the cathode is to be used, in a future 
experiment for \Bapp\ tagging, as envisioned by NEXT-BOLD. 

The evolution of this international context is still unclear. Neither PANDAX-III nor AXEL have so far published results based on large prototypes such as NEXT-DEMO or NEXT-DBDM (in fact most of the quantitative data was produced by the prototype NEXT-MM), and no radiopure demonstrator such as \NEW\ has yet been operated. There are clearly opportunities for convergence, in particular between NEXT and AXEL collaborations, and the eventual possibility of two experiments deploying HPXe TPCs, one operating in Europe (LSC, LNGS) or Canada (Snowlab) and another in China (JinPing) may, eventually turn into an opportunity for the technology.    
 
\subsubsection{nEXO}

The next-generation Enriched Xenon Observatory nEXO is designed to optimize the unique features of a massive  monolithic and  homogeneous detector using 5-tons of isotopically enriched liquid xenon. The implementation of LXe-based  time projection chambers (TPCs) has already been successfully demonstrated on a smaller scale by the predecessor EXO-200.

As in EXO-200, the nEXO TPC is a LXe single-phase device resulting in fewer components and lower background. The primary goal is the optimization of the energy resolution near the Q-value.
The nEXO TPC is designed to read out both ionization and scintillation light, in order to exploit the anti-correlation between these two channels and to obtain the best possible energy resolution.  Charge collection will be achieved at the top of the TPC, using gold strips deposited on silica ``tiles''. Scintillation readout will be obtained with VUV-sensitive Silicon Photomultipliers (SiPMs) installed behind the field-shaping rings. This location will allow for larger coverage compared to the case of EXO-200, where photodetectors were installed behind the anode grids.   

The LXe TPC is at the center of different active and passive shielding layers, each containing components made of materials which are progressively lower in radioactive contamination the deeper they are in the detector. As successfully done in EXO-200, the innermost shielding layer will consist of a bath of HFE-7000, at least 76~cm thick in all directions. With a density of 1.8~g/cm$^3$ at 170~K, this fluid is an efficient $\gamma$-ray shield and is one of the most radiopure materials identified by the EXO-200 and nEXO screening campaigns. The large cold mass, mainly composed of HFE-7000, further provides substantial thermal inertia, making the cryogenic system intrinsically stable. The cryostat is composed of two nested vessels, separated by vacuum insulation. Substantial infrastructure is required to cool the xenon and the HFE-7000 as well as for recirculation and purification of  the xenon.

nEXOs sensitivity reach is based on assumptions on the detector and analysis performance, and on using only measured radioassay inputs to build the background model. 
nEXO’s expected median sensitivity on the \Tonu\ at 90\% C.L. reaches \SI{9.2E+27}{\yr} in a 10 year run. A 3$\sigma$ discovery potential of \SI{5.7E+27}{\yr} is predicted for the same live time \cite{nexo-sensitivity-Albert:2018bc}.

\subsubsection{Loaded LSc}

Large volume liquid scintillator detectors loaded with a DBD isotope represent a cost-efficient way of scaling up an experiment to large isotope masses. Deep purification and highly efficient vetoing techniques developed for such detectors searching for solar neutrinos (e.g. Borexino) allow a very low background index to be reached. Two collaborations are pursuing this approach: KamLAND-Zen and SNO+. Both experiments are reusing the existing detector infrastructure from previous reactor and solar neutrino studies. This approach has a relatively poor energy resolution and somewhat limited particle identification. This is however at least partly compensated by very large isotope masses achievable with this technology. 

The KamLAND-Zen detector is located in the Kamioka mine in Japan. The detector contains 13 tons of Xe-loaded liquid scintillator suspended in a transparent nylon-based inner ballon surrounded by 1 kton of liquid scintillator.  KamLAND-Zen currently holds the most sensitive constraint on the effective Majorana neutrino mass. Using 380~kg 
of $^{136}$Xe they established a lower limit on 
the DBD$0\nu$ half-life of $> 1.07 \times 10^{26}$ yr corresponding to 
$| m_{\beta\beta} |  < 0.061$--$0.165$ meV. The experiment is currently taking data with an increased loading 
of 750~kg if $^{136}$Xe aiming to reach a sensitivity of $T_{1/2}^{0\nu}  > 4.6 \times 10^{26}$ yr. 

SNO+ will repurpose the infrastructure used for the SNO experiment in the SNOLAB laboratory near Sudbury (Canada). 
The detector will be filled with 800 tons of linear alkyl benzene liquid scintillator loaded with $^{130}$Te. A high natural 
isotopic abundance (33.8\%) provides the opportunity to avoid the enrichment process. The tellurium loading is expected to start at the end of 2019. In the first phase of the experiment a loading of 0.5\% is expected. The expected sensitivity for this phase is  $T_{1/2}^{0\nu}  > 1.9 \times 10^{26}$ yr after 5 years of data taking. 

KamLAND-Zen and SNO+ collaborations are planning future phases of the experiment to increase their sensitivity. Both experiments plan to increase the photocathode coverage and drastically improve the scintillator light collection to a level of 1000~p.e./MeV which would allow them to reach an energy resolution of 4-5\% (FWHM) at the $Q_{\beta\beta}$ value of 
$^{130}$Te and $^{136}$Xe. In addition the isotope loading will be increased with a key challenge to maintain the transparency and light yield of the scintillator and the uniformity and stability of the loading. 

\subsubsection{SuperNEMO}
The SuperNEMO collaboration pursues a very different approach that involves a full topological reconstruction 
of individual electrons emitted in the DBD decay. A DBD source in the form of a thin foil is surrounded by a low-density tracker and a fast calorimeter. The unique features of this approach are the ability to study almost any 
DBD isotope and reconstruction of the event topology which produces a "smoking gun" evidence for the process and may allow the underlying physics mechanism to be disentangled. Its technology is based on the NEMO-3 experiment which was running at the Modane Underground Laboratory (LSM) in the Frejus tunnel in 2003--2011 and produced the most accurate DBD$2\nu$ half-life measurements for 7 different isotopes and a competitive DBD$0\nu$ constraint for 
$^{100}$Mo. This technology is uniquely suitable for precision studies of the DBD$2\nu$ decay providing important experimental input to nuclear models for NME calculations and to possible quenching of $g_A$, as well as for more exotic models of DBD$0\nu$ (e.g. with a Majoron emission) and other new physics scenarios (Lorentz violation, bosonic neutrino etc). 

The first SuperNEMO  module will start running in 2019 at the LSM underground laboratory as a technology demonstrator. The full version of SuperNEMO with 100~kg of $^{82}$Se and 20 demonstrator-like modules is expected to reach a sensitivity 0.05--0.1eV. The experiment has a large footprint and currently does not have a hosting lab for its full version. The current collaboration strategy is to fully exploit the physics with the first demonstrator module including exotic physics scenarios mentioned above and understand the background levels reachable with this technology. In the event of a $0\nu\beta\beta$ discovery corresponding to $| m_{\beta\beta} | \geq$ 0.05~eV with any other technology SuperNEMO will be in an excellent position to confirm the observation with different isotopes and a good potential to disentangle the underlying physics mechanism using its ability to measure individual electron tracks (e.g. the V+A currents contribution to $0\nu\beta\beta$).   

\subsubsection{COBRA}
In the general calorimetric approach, the detector has to contain the double beta decay isotopes: COBRA (Cadmium-zinc-telluride 0-neutrino double Beta Research Apparatus) experiment exploit the semiconductor CdZnTe detector technology that, at the same time, contains nine double beta decay isotopes: five decays ${\beta^-\beta^-}$ and four ${\beta^+\beta^+}$. In particular, $^{116}$Cd shows a high transition energy, 2813 keV: the neutrinoless double beta decay peak will lie in a relatively low background region of the energy spectrum. Another important isotope that could be measured is $^{130}$Te with a relatively high transition energy, 2527 keV, but having the largest isotopic abundances, 34\%, for a double beta decay active isotope. CdZnTe is a semiconductor detector that operate at room temperatures. It is normally used as radiation detector for its high energy resolution that help in the identification of the experimental background sources.

COBRA detector geometry is focused on an array of multiple CdZnTe detector crystals able to exploit the advantages of two basic detector concepts: the detector crystals are both source and detector allowing a high detector efficiency; using a detector array it could be possible to disentangle multiple hits, due typically to gamma rays interaction, to single hit, strictly related to the neutrinoless double beta decay. The experiment was designed as an array of 1600x1 cm$^{3}$ CdZnTe cubes for a total detector mass of around 10 kg. In a recent redesign, the single crystal was modified to a much larger one with a size of 2x2x1.5 cm$^{3}$. This new detector configuration will also help in reducing an important background contribution that is produced by radioactive decay on detector surfaces. In the new configuration a larger detector mass could be possible and a more efficient background rejection is achievable. 

A detector demonstrator of the COBRA experiment is in data taking at Gran Sasso Underground Laboratory with two detector configurations: the original one with crystals having 1 cm$^{3}$ volume and in the XDEM (eXtended demonstrator) with the new large single detector. The current setup has an array of 64 cubic CdZnTe detector crystals of 1 cm$^{3}$ for a total mass close to 400 g and 9 XDEM CdZnTe detector crystals for an additional mass of around 300 g. The experimental background level in the $^{116}$Cd ROI is of the order of 0.1 counts/(keV kg year). COBRA demonstrator is also measuring $^{113}$Cd beta decay: very suppressed single beta decays gives important information on the nuclear decay mechanisms and will help in determine the nuclear matrix elements.

\subsection{Comparisons between different experimental efforts}

Although different in technology, scale and reach, we compare here the three next generation experiments we have focussed on, highlighting their key features, their complementarity and their sensitivities.

\Lthou~ builds on the \GERDA~ and \MJD~ experience and on the  R\&D carried out for \Ltwo . The baseline technology of enriched HPGe detectors and the liquid argon instrumentation is well established. Improvements in detector performance and background reduction are well advanced and the background reduction goals with respect to the state-of-the-art experiment \GERDA~  corresponds to about a factor three for \Ltwo . For \Lthou~ one additional order of magnitude is required to operate background-free and exploit the full discovery potential. Reduction of surface decays of $^{42}$K, progeny of $^{42}$Ar, and further improvement of the purity of small components close to the detectors are critical to reach this goal. The \LEG~ collaboration pursues a rigorous staging of the experiment to expedite physics results.

The technology of Li$_2$MoO$_4$ scintillating bolometers – to be used to study the promising isotope $^{100}$Mo –  is fully developed, including purification of enriched material, crystal growth, and construction and validation of the detectors. The prototypes fabricated so far have shown an outstanding behavior in terms of energy resolution, internal radiopurity and $\alpha$-particle rejection factor, confirmed at a medium scale by the CUPID-Mo experiment in LSM. 
On the basis of these results, the CUORE collaboration decided in May 2018 to adopt the Li$_2^{100}$MoO$_4$ technology as a baseline --- keeping TeO$_2$ as a backup --- for the proposed future experiment CUPID, to be installed in the existing CUORE cryostat. The CUPID collaboration is in formation and the Conceptual Design Report is currently being finalized. The full CUPID experiment implies of course a substantial investment in enrichment and crystallization. However, CUPID is cost-effective and can count on an existing infrastructure, although some upgrades are envisioned. The available background model --- taking advantage of the high Q-value of $^{100}$Mo placed beyond the $\gamma$ environmental radioactivity --- predicts a background index of the order of $1 \times 10^{-4}$\ckky. It is to remark that this model exploits real data collected by CUORE directly inside the infrastructure that will be used by CUPID.
In terms of time schedule, about 3 years from now are required to fix definitely the CUPID detector structure and to study the cryostat upgrades. In parallel, enrichment and crystallization can be completed in 5 years, as well as the detector construction and assembly, that can start as soon as the final structure is fixed. The requested improvements of the cryogenics can be implemented in the year that follows the completion of the five-year CUORE scientific program. Therefore, CUPID could be ready for commissioning and data taking in 6 years from now. 

The readiness of the HPXe technology has been demonstrated by the NEXT-White detector, currently operating at the LSC, which has validated the excellent energy resolution \cite{Renner:2019pfe}, and the discrimination of the topological signature \cite{Ferrario:2019kwg}. Furthermore, the NEXT-White background model has been established, showing a good agreement with expectations \cite{NewRadiogenic}. NEXT-White will take data until the commissioning of NEXT-100 (in 2020). A measurement of the DBD2$\nu$ mode as well as a search for DBD0$\nu$ events aimed to quantify the technology background index is under way.

NEXT-100 will start operations in 2021. While the physics reach of the experiment is competitive with existing efforts, the detector can also be considered as a prototype of the next-generation experiments. The expected background index is very low ($4 \times 10^{-4}$\ckky), and is expected to be dominated by the PMTs. Operation of NEXT-100 combined with R\&D should be able to assess precisely the expected background index for the first next-generation module, expected to be NEXT-HD.

On the other hand, the possibility of implementing barium tagging in NEXT appears today as a promising possibility \cite{McDonald:2017izm}. Provided that barium tagging can be realized with high efficiency, a $3 \sigma $ discovery sensitivity at the $10^{28}$~yr is conceivable. R\&D over the next few years should assess the feasibility of this disruptive approach, that would lead to a virtually ``background-free'', NEXT-BOLD detector.

\subsubsection{Discovery potential}

We provide here the sensitivities in terms of half-lives and $m_{\beta \beta}$ reach of the experiments. This analysis follows closely the strategy developed in Ref.~\cite{Agostini:2017jim}. The primary experimental signature for \nubb\ decay  is
a mono-energetic peak in the measured energy spectrum at the $Q$-value of the decay,
produced when the two electrons emitted in the process are fully absorbed in the
detector active volume.
While in many detectors additional analysis handles are available to distinguish
signal from background, energy is the one observable that is both necessary and sufficient
for discovery, and so the sensitivity of a DBD$0\nu$ decay experiment is
driven by Poisson statistics for events near the $Q$-value. It can thus be approximated with a heuristic counting analysis, where there are just two parameters of interest: the
``sensitive exposure'' (\senexp) and the ``sensitive background'' (\senbkg). 
\senexp\ is given by the product of active isotope mass 
and live time, corrected by the active fiducial volume, the signal detection
efficiency, and the probability for a \nubb\ decay event to fall
in the energy region of interest (\senroi) in which the experiment is sensitive to the
signal. \senbkg\ is the number of background events in the \senroi\ after all analysis
cuts divided by \senexp.
The number of signal and background counts in the final spectrum
is then given by:
\begin{equation}
   N_{\nubb} = \dfrac{\ln 2 \cdot N_{A} \cdot \senexp}{ m_a \cdot\hl} \qquad
   \text{and} \qquad N_{bkg}= \senbkg\cdot \senexp ~,
\end{equation}
where $N_A$ is Avogadro's number,  $m_a$ is the
molar mass of the target isotope, and \hl\ is the half-life of the decay.

Following the proposal of Ref.~\cite{Agostini:2017jim}, 
the sensitivity of an experiment to discover a signal is defined as the
value of \hl\ or $|m_{\beta \beta}|$ needed to exclude the no-signal hypothesis with a median
significance of 99.7\% confidence level. 
The previous statistical statement can be rephrased in words more accessible to the
physics community saying that the \hl\ sensitivity is the minimal
signal strength that an experiment will measure in 50\% of the cases with a
significance of at least 3$\sigma$. Such a value is hence converted in terms of
\mbb\ using the collection of matrix elements discussed in Sec. 3.
Further information on how the calculation is performed can be found in the Appendix of Ref.~\cite{Agostini:2017jim}.

The efficiency considered in the calculation are:
the fraction of isotope mass used for analysis 
(accounting for dead volumes in solid detectors and fiducial volume cuts
in liquid and gaseous detectors), 
the analysis cuts meant to enhance the rate of signal-to-background events,
and the fraction of fully-contained DBD$0\nu$ decay events with energy
reconstructed in the \senroi. 
The choice of optimal \senroi\ depends on the background rate, its energy distribution, and
the energy resolution ($\sigma$) of the Gaussian peak expected from the signal.
Experiments with an excellent energy resolution ($\sigma<1\%$) have a \senroi\
centered at the $Q$-value with a width depending on the background rate.
For experiments with poorer energy resolution, the background
due to two-neutrino double-$\beta$ decay is significant up to the
$Q$-value. These experiments have an asymmetric optimal \senroi\ covering
primarily the upper half of the Gaussian signal. 

\begin{figure}[ht!]
    \centering
    \includegraphics[width=0.6\textwidth]{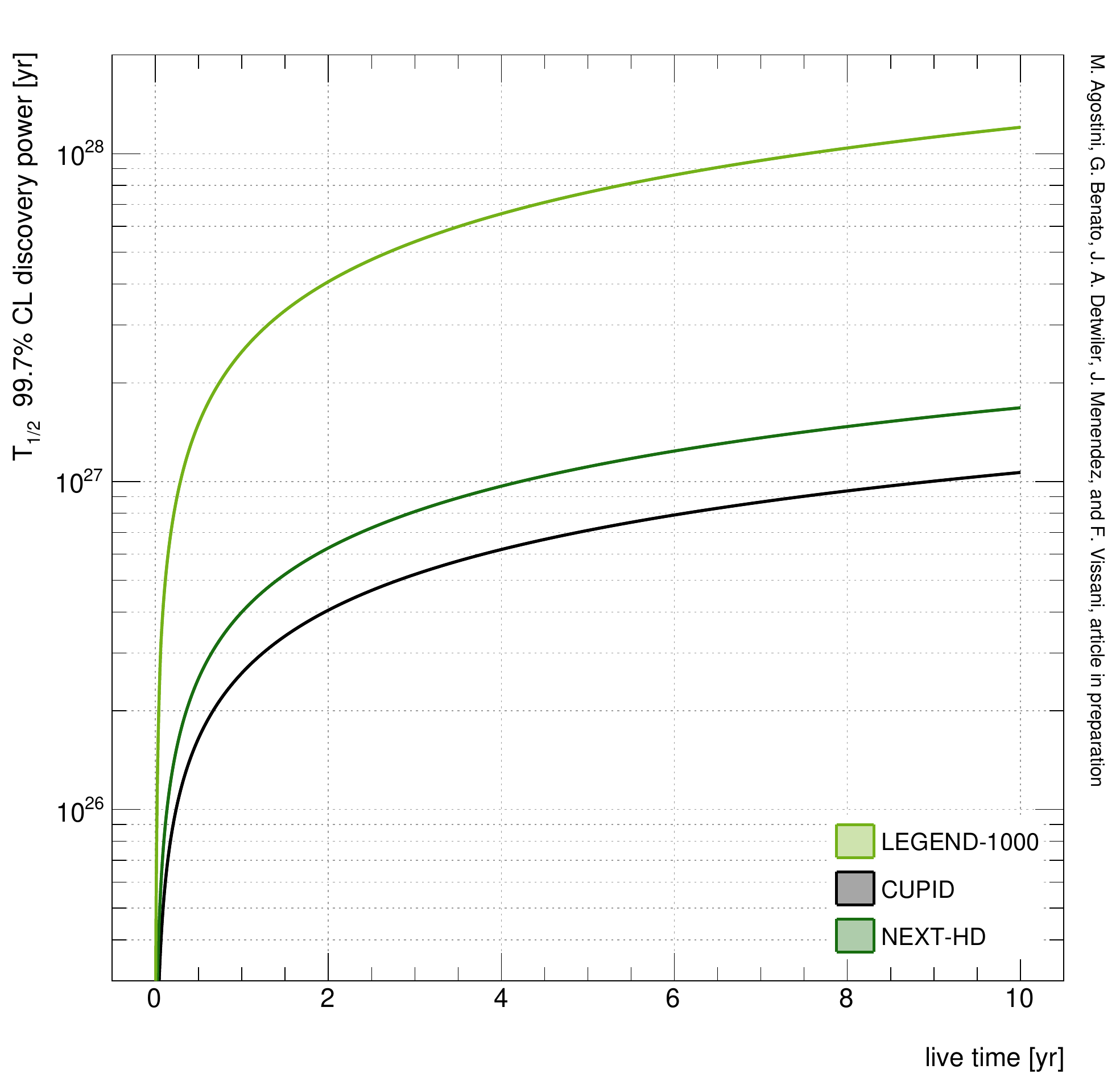}
    \caption{Discovery sensitivity for $T^{0\nu}_{1/2}$ for \Lthou, CUPID and NEXT-HD  at 3$\sigma$ as a function of live time. Provided by M. Agostini, based on Ref.~\cite{Agostini:2017jim}.}
    \label{fig:T12}
\end{figure}

\begin{figure}[ht!]
    \centering
    \includegraphics[width=0.8\textwidth]{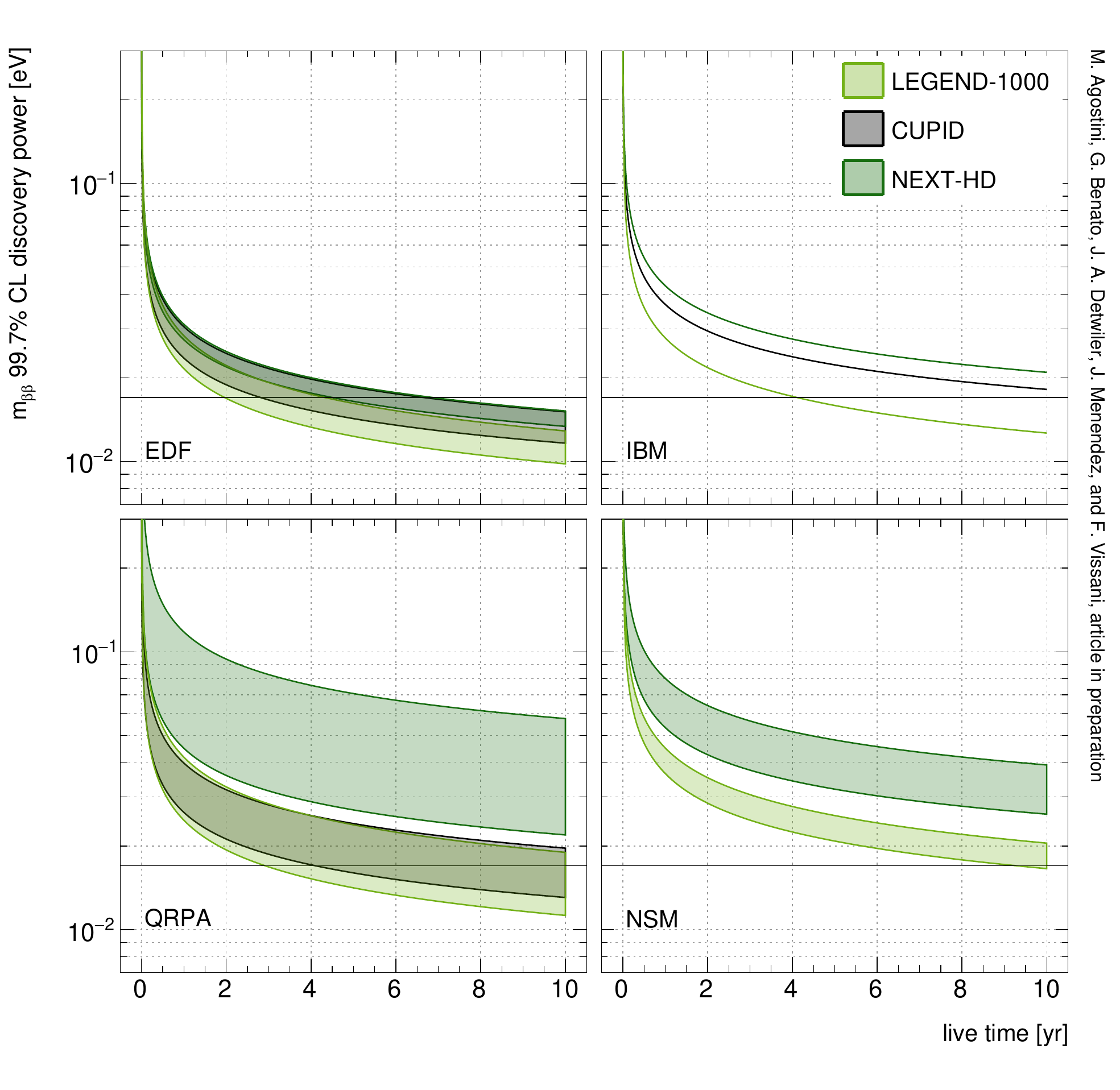}
    \caption{Discovery sensitivity for $m_{\beta\beta}$ at 3$\sigma$ for \Lthou, CUPID and NEXT-HD as a function of live time. The current spread of NMEs for each nucleus has been included. No ISM calculation is available for $^{100}$Mo at the moment.  Provided by M. Agostini, based on Ref.~\cite{Agostini:2017jim}.}
    \label{fig:mbb}
\end{figure}

In cases where energy spectral fits and position non-uniformity enter non-trivially into the sensitivity (as e.g. in nEXO), the parameters are tuned to match the collaboration's
stated sensitivity. It should be emphasized that the projected sensitivities are
not meant to directly compare one experiment to another:
many experiments are under rapid development, and the
parameters publicly available during the snapshot of time in which this document
was prepared will often poorly characterize their ultimate reach.
In addition such a heuristic analysis does not consider the very important issue
that the background expectations in the \senroi\ are affected by systematic
uncertainties due to the background modeling. 
We report the median 3$\sigma$ discovery sensitivities for $T_{1/2}$ and $|m_{\beta \beta}|$ in Figs.~\ref{fig:T12} and \ref{fig:mbb}, respectively.

\begin{table}[hbth]
\small
   \centering
   \begin{tabular}{|c|c|c|c|c|c|c|c|c|c|}
   \hline 
       Current & Iso & $M_{iso}$ &  $\sigma$ & ROI & $\epsilon_{sig}$ & $\mathcal{E}$ & $\mathcal{B}_\mathrm{ROI}$ & \multicolumn{2}{c|}{Sens./Lim. (90\%C.L.)} \\
       experiments & & [kg] &  [keV] & [$\sigma$] 
       & [\%]& $\left[ \frac{\mathrm{kg}_{iso} \mathrm{yr}}{\mathrm{yr}} \right]$ &  $\left[ \frac{\mathrm{cts}}{ \mathrm{kg}_{iso} \mathrm{yr}} \right]$ & $T_{1/2}$ [yr] & $|m_{\beta \beta}|$ [meV] \\
       \hline 
       GERDA & $^{76}$Ge & 31 & 1.4 & (-2.0, + 2.0) & 60 & 19 & $6\cdot10^{-3}$  & $1.1/0.9 \cdot10^{26}$ & 102-213 \\
       CUORE & $^{130}$Te & 206 & 3.4 & $-1,4 , + 1,4$ & 67 & 138 & $6.7\cdot10^{-1}$ & $2.3\cdot10^{25}$ & 90-420 \\
       \hline \hline 
    Current & Iso & $M_{iso}$ &  $\sigma$ & ROI & $\epsilon_{sig}$ & $\mathcal{E}$ & $\mathcal{B}_\mathrm{ROI}$ & & \\
       demonstrator & & [kg] &  [keV] & [$\sigma$] 
       & [\%]& $\left[ \frac{\mathrm{kg}_{iso} \mathrm{yr}}{\mathrm{yr}} \right]$ &  $\left[ \frac{\mathrm{cts}}{ \mathrm{kg}_{iso} \mathrm{yr}} \right]$ &  &  \\
       \hline 
       CUPID-0 & $^{82}$Se & 4,65 & 8.5 & $-2.0 , + 2.0$ & 70 & 3.3 & $2.2\cdot10^{-1}$ & $3.5\cdot10^{24}$ & 311-638 \\
       CUPID-Mo & $^{100}$Mo & 2.26 & 2.3 & $-2.0 , + 2.0$ & 64 & 1.44 & - & - & - \\
     NEXT-White    & $^{136}$Xe & 91      & 10  &   $-1.0 , + 1.9$ & 26 & -   & - & - & - \\
       \hline \hline 
     Funded & Iso & $M_{iso}$ &  $\sigma$ & ROI & $\epsilon_{sig}$ & $\mathcal{E}$ & $\mathcal{B}_\mathrm{ROI}$ & \multicolumn{2}{c|}{3$\sigma \ \mathrm{disc. \ sens.}$}  \\
       experiments & & [kg] &  [keV] & [$\sigma$] 
       & [\%]& $\left[ \frac{\mathrm{kg}_{iso} \mathrm{yr}}{\mathrm{yr}} \right]$ &  $\left[ \frac{\mathrm{cts}}{ \mathrm{kg}_{iso} \mathrm{yr}} \right]$ & $T_{1/2}$ [yr] & $m_{\beta \beta}$ [meV] \\
       \hline 
      LEGEND-200  & $^{76}$Ge  & 177  &     1.1 &   $-2.0 , + 2.0$ & 70 & 123  & $1\cdot10^{-3}$& $9.4\cdot10^{26}$ & 35--73  \\     
      NEXT-100    & $^{136}$Xe & 87      & 10.4  &   $-1.0 , + 1.8$ & 26 & 23   & $4\cdot10^{-2}$& $7.0\cdot10^{25}$ &~65--281 \\      
       \hline \hline 
     Future & Iso & $M_{iso}$ &  $\sigma$ & ROI & $\epsilon_{sig}$ & $\mathcal{E}$ & $\mathcal{B}_\mathrm{ROI}$ & \multicolumn{2}{c|}{3$\sigma \ \mathrm{disc. \ sens.}$}  \\
       experiments & & [kg] &  [keV] & [$\sigma$] 
       & [\%]& $\left[ \frac{\mathrm{kg}_{iso} \mathrm{yr}}{\mathrm{yr}} \right]$ &  $\left[ \frac{\mathrm{cts}}{ \mathrm{kg}_{iso} \mathrm{yr}} \right]$ & $T_{1/2}$ [yr] & $m_{\beta \beta}$ [meV] \\
       \hline 
       LEGEND-1000 & $^{76}$Ge  & 883     & 1.1 &   $-2.0 , + 2.0$ & 70 & 614  & $7\cdot10^{-5}$& $1.2\cdot10^{28}$ & 10--20  \\
       \hline
       CUPID       & $^{100}$Mo & 253   & 2.1 &   $-2.0 , + 2.0$ & 68 & 172  & $2\cdot10^{-3}$& $1.1\cdot10^{27}$ & 12--20  \\
       \hline
       NEXT-HD     & $^{136}$Xe & 991     & 7.7 &   $-1.3 , + 2.5$ & 32 & 317  & $9\cdot10^{-4}$& $1.7\cdot10^{27}$ & 13--57 \\
       \hline
\hline
   \end{tabular}
   \caption{Experimental parameters of European next-generation experiments. Iso refers to the isotope used and $\mathrm{M}_{iso}$ to its mass. $\sigma$ is the energy resolution to the standard deviation. ROI refers to the Region of interest given in units of $\sigma$ from the Q-value of the decay. 
   $\epsilon_{sig}$ is the total signal detection efficiency. 
   The sensitive exposure $\mathcal{E}$ and background $\mathcal{B}_{\mathrm{ROI}}$ in the ROI are normalized to 1 yr of live time. For the current experiments the achieved 90\% C.L. limits are given together with the experimental sensitivity in case of no signal. The published limits have been derived from a full likely-hood analysis and the ROI is given only for illustration. Instead for the future experiments, the the median 3$\sigma$ discovery sensitivities for  
   $T_{1/2}$ and $|m_{\beta \beta}|$ are reported  assuming 10 years of live time (5 years for LEGEND-200), including the different NME calculations. Provided by M. Agostini, based on Ref.~\cite{Agostini:2017jim}.}
   \label{tab:my_label}
\end{table}

For completeness and comparison, we give the key parameters for the US-led nEXO experiment and the Japanes led KamLAND2-Zen experiments. The $T_{1/2}$  3$\sigma $ discovery potential of nEXO corresponds to $5.3\cdot10^{27}$~yr or $8-32$~meV ($|m_{\beta\beta}|$) deploying 4605 kg of $^{136}$Xe. The Japanese-led KamLAND2-Zen experiment plans to dissolve 1000~kg of $^{136}$Xe in a liquid scintillator and estimates a 3$\sigma $ discovery sensitivity of $1.2\cdot10^{27}$~yr corresponding to $23-49$~meV ($|m_{\beta\beta}|$).

\subsection{A path towards experiments with meV sensitivity} 

If a positive signal is not observed in the next generation DBD0$\nu$ experiments there is a compelling motivation to push the sensitivity further with a clear target of reaching a $\sim$meV sensitivity which would cover a large part of the parameter space corresponding to the normal neutrino mass ordering. This is an enormously ambitious task but an R\&D program for a practical experiment capable of reaching such sensitivity should be encouraged and pursued in parallel with construction and operation of next generation DBD0$\nu$ experiments. 

In order to start probing the neutrino mass normal ordering one has to improve the sensitivity to the half-life by at least an order of magnitude compared to the next generation DBD0$\nu$ experiments. There are two critically important challenges that need to be addressed to achieve such sensitivity levels: 

\begin{enumerate}
\item The detectors will need to host an order of magnitude larger mass of $\beta\beta$ isotope, at the 10~ton level. In general, an exposure at a level of $\sim$100 ton$\times$yr will be required. 
\item The background index must be reduced by a further order of magnitude to achieve a level of $< 0.01 \mathrm{cnts}/(FWHM \cdot t \cdot yr)$.  
\end{enumerate}

A breakthrough in enrichment technologies would greatly help addressing the first challenge. An increase in the productivity of the enrichment process and most importantly a reduction in cost should be the key avenues to be explored. Developing a dedicated international facility for stable isotope enrichment can be an efficient way of addressing this issue. It is worth noting that there are significant synergies with industrial and knowledge exchange programs where enrichment of stable isotopes is required. Isotopes of $^{130}$Te and $^{136}$Xe have an interesting potential for 10~ton detectors. Due to its high natural abundance $^{130}$Te is probably the only isotope that could be used without enrichment while the enrichment of $^{136}$Xe is cheaper compared to other isotopes since the chemistry process required to prepare a gas suitable for centrifuges is not needed. The latter is partly counterbalanced by a high cost of xenon. 

The second challenge will require a breakthrough in ultra-low background technologies. Although extremely challenging the technologies  pursued in Europe (HPGe, scintillating bolometers and HPXe-TPC) do not have a "no-go theorem" that can stop them from further improving on the backgrounds. Due to their exquisite energy resolution and the intrinsic purity of the crystals HPGe and bolometer technologies will be able to reach a background free regime from the detectors themselves. Moreover, the HPGe detector technology effectively eliminates 
the "ultimate" DBD$2\nu$ background. The key challenges for  both HPGe and bolometers will be to remove or drastically reduce the background in the vicinity of the crystals due to front-end electronics hardware and support structures, as well as to come up with a "smart" and extremely efficient active veto system. The HPXe-TPC technology has a very challenging but equally powerful "trick up the sleeve" $-$ the barium tagging as described in the NEXT-BOLD section. If successful, this has the potential of eliminating all backgrounds except DBD$2\nu$ and a good energy resolution offered by HPXe-TPC and a long half-life of $^{136}$Xe DBD$2\nu$ decay could allow the experiment to reach the required backgrounds. 

An R\&D on the undergraduate infrastructure requirements commensurate with the background targets described above will have to be carried out. The questions of the depth and external background suppression (neutron and gamma flux, radon levels) as well as production of detector components underground to avoid cosmic activation will need to be addressed. 

In summary, an ambitious R\&D program should be pursued in order to identify a viable approach to reach a half-life sensitivity to DBD0$\nu$ exceeding 10$^{29}$ yr that will allow exploring the normal ordering of neutrino masses at $\sim$~meV scale with a tremendous potential of discovering lepton number violation.

\subsection{Broader Physics Program}
Although DBD0$\nu$ decay is the main objective, a number of other processes is open to experimental investigation and the anticipated low background rate promises competitive sensitivities for many of them. The accessible processes include alternative modes of double beta decay as well as more exotic processes predicted by some extensions of the Standard Model. The experimental investigation of the validity limits of fundamental principles like charge conservation of CPT/Lorentz invariance deserves particular attention since most of out theoretical construction is based on them.

Finally, the search for Dark Matter candidates, originated from the first experiments on DBD0$\nu$ decay, is still one of the most appealing objectives.

\subsubsection{\LEG~ and Dark Matter}
Dedicated dark matter experiments based on germanium technology optimize the energy threshold to boost their sensitivity for low-mass WIMP dark matter particles. This is usually achieved by operating germanium crystals as cryogenic calorimeters to increase the sensitivity for nuclear recoils, as performed in Edelweiss and the future SuperCDMS experiments. HPGe detectors  employed in \Gerda~ and \MJD~ are operated as diodes and can reach sub-keV thresholds for electron recoils. Competitive limits in the few ~GeV mass range have recently been provided by the CEDEX collaboration \cite{Jiang:2018pic}. \Gerda~ and in the future,  \Ltwo~ operate the HPGe in natural argon. Backgrounds from $^{39}$Ar beta decays in the surrounding argon dominate the count rate at low energies and thus limit the sensitivity for WIMP dark matter searches. \Lthou~ instead plans to operate the HPGe detectors in argon from underground gas wells, which is depleted not only in  $^{42}$Ar but also in $^{39}$Ar. Therefore, \Lthou~ will extend the physics program towards low-energies, including dark matter search. Other BSM  physics with competitive or even leading sensitivities, as performed recently by the \Gerda~ and \MJD~ experiments, include the searches for bosonic dark matter \cite{Abgrall:2016tnn},  baryon decay \cite{Alvis:2018pne}, for solar axions, Pauli exclusion principle violation, and electron decay \cite{Abgrall:2016tnn}, or for Majoron emission \cite{Agostini:2015nwa}. 

\subsubsection{CUPID and Dark Matter}
Cryogenic bolometers have a long history in direct dark matter search. The classical and most important  class  of  potential  dark  matter  candidates searched for in the last two decades are Weakly Interacting Massive Particles (WIMPs). Currently the experiments with the highest sensitivity to WIMPs in the mass-range above 5 GeV/c$^{2}$ rely on ton-scale LXe TPC as they provide a very low-background combined with a large target mass. 

The cryogenic bolometer experiments  using Ge-crystals (EDELWEISS, (Super)CDMS)  and scintillating  CaWO$_{4}$-crystals (CRESST) cannot compete in the search for medium to high-mass WIMPs. However, thanks to their excellent energy resolution and low energy threshold for nuclear recoils (below 1 keV) they lead the field in the low-mass dark matter search ($<$0.5 GeV/c$^{2}$). CRESST-III achieved with present generation of such detectors in combination with employing a light target element (oxygen) sensitivity to dark matter particle masses in the sub-Gev regime.

A next-generation CUPID experiment consisting of a ton-scale array of Li$_2$MoO$_4$ crystals operated as scintillating bolometers fulfills the requirements for a dark matter search for sensitivity in mid-to-low mass regime as it provides:  light target elements (Li, O), discrimination between nuclear recoils events (as expected for dark matter interactions) and $\beta$ / $\gamma$-events, extremely low background and a low energy threshold. Relying on a performance for CUPID similar to the one achieved in a measurement in the R\&D facility of CUPID at LNGS a competitive dark matter sensitivity may be reached in the mass-range of 1-5 GeV/c$^{2}$.

\subsubsection{NEXT and Dark Matter}
Concerning dark matter searches, a gas detector may presents a number of advantages over the liquid phase, including the potential ability to reject few-electron background events at low energies and the ability to operate with significant amounts of a light gas (such as He or Ne) to increase sensitivity at lower dark matter masses ($m_{\chi}\sim$ 1 GeV).  Therefore a ton-scale NEXT module could search for dark matter using two different approaches: a search for dark matter-electron interactions at the MeV mass scale, or a search for dark matter-nucleon interactions at the several-GeV mass scale. In this document we will only highlight, the case of MeV-range dark matter, where gas detectors may be complementary to LXe experiments. 

Light dark matter (LDM), or dark matter particles with sub-GeV masses that are too small to create detectable nuclear recoil signals in the majority of today's WIMP search experiments, present a potential front for NEXT.  Dark matter with mass in the 10-1000 MeV range could create detectable signals via interactions with \textit{electrons} in noble gases.  Such signals would be small (one to several ionization electrons), but could be as numerous as thousands of events per year in a kg-scale background-free detector with single-electron sensitivity.  The measurement of such small signals requires a sensitive detector and thorough understanding and control over backgrounds.  Several WIMP search experiments based on liquid Ar \cite{DarkSide_2018} and Xe \cite{Essig_XENON10, Essig_XENON100} have already performed analyses of low-energy events and set limits on the dark matter-electron (DM-e$^{-}$) interaction cross section $\bar{\sigma}_{e}$.  However, these limits have been quite conservative due to a lack of full understanding of few-electron background events.  It seems possible, therefore, that a significant amount of unexplored parameter space could be reached even with a modest exposure. In particular, one of the most prominent backgrounds due to field emission of electrons from high-voltage cathode grids, can be uniquely controlled in a gaseous active medium by observing an initial flash of photons produced in the high fields near the cathode upon field emission. Clearly, the potential of gaseous detectors for dark matter searches deserve serious study. 

\subsubsection{Other double beta decay processes}
The best double beta decay experimental sensitivity is generally for the transition to the ground state of the daughter nucleus. However double beta decay may occur (in both 2$\nu$ and 0$\nu$ modes) also to an excited state of the daughter nucleus. 
In the case of neutrinoless double beta decay these transitions can disclose the exotic mechanisms (eg RH currents) which mediate the decay,~\cite{TOMODA2000245,SIMKOVIC2002201} while for two neutrino double beta decay they can provide unique insight to the details of the mechanisms responsible for the nuclear transition~\cite{SUHONEN1998124}.
From the experimental point of view, most of the interest is motivated by the fact that in a close packed array, like CUPID, the strong signature  provided by the simultaneous detection of one or two gammas can lead to an almost background-free search. 
In this respect, the transitions to 0+ states are favoured while states with larger spin (e.g. 2+) are generally suppressed by angular momentum conservation.

$\beta^+\beta^+$, $\epsilon\beta^+$ and $\epsilon\epsilon$ modes are generally less appealing because of the lower available energy. However they can only be mediated by peculiar mechanisms and can therefore provide unique information on the decay details~\cite{VERGADOS1983109,IACHELLO2014064319}.

Exotic neutrinoless double beta decays characterized by the emission of a massless Goldstone boson, called Majoron, are predicted by some theoretical models~\cite{GELMINI1981411}. The precise measurements of the Z invisible width at LEP, has greatly disfavoured the original Majoron triplet and pure doublet. However, several new  models have been developed ~\cite{BURGESS19945925,BAMERT199525}. All these models predict different (continuous) spectral shapes for the sum energies of the emitted electrons, which extend from zero to the transition energy Q$_{\beta\beta}$:
\begin{equation}
\frac{dN}{dT} \sim (Q_{\beta\beta}-T)^n
\end{equation}
where T is the electron summed kinetic energy and the spectral index n depends on the decay details. Single Majoron emissions are characterized by n=1-3 , while double Majoron decays can have either n = 3 or n = 7. The precise measurement of n allows to discriminate between the processes.
As for any process characterized by continuous spectra, the experimental sensitivity is mainly limited by the background contributions and the detector mass~\cite{ELLIOTT19871649,ARNOLD2006483,KZENMAJORON2012}.

\subsubsection{Violations of Fundamental Principles}
The decay of an atomic electron is probably the most sensitive test of electric charge conservation. 
Charge non conservation (CNC) can be obtained by including additional interactions of leptons and photons which lead to the decay of the electron: $e \to \gamma \nu$ or $e \to \nu_e\nu_X\bar{\nu}_X$. These modes  conserve all known quantities apart from electric charge.
An additional possibility is connected with CNC involving interactions with nucleons. Discussions of CNC in the context of gauge theories can be found in a number of BSM gauge models~\cite{VOLOSHIN1978145,OKUN1978597,IGNATIEV1979315}.

While the signature of the neutrino mode is quite poor, the coincidence between the decay gamma and the atomic de-excitation X-rays can give rise to interesting topological configurations which can help to lower the background contributions. 
The most stringent limits on CNC have been obtained as side results in experiments characterized by large masses and very low backgrounds~\cite{DAMA2000117301,BACK200229}.
Indeed, the large detection efficiency, low threshold and excellent energy resolution expected for the bext generation double beta decay experiments are crucial to detect the low energy de-excitation X-rays or Auger electrons and, associated to the ton-size scale of the experiment, anticipate competitive results. 

Lorentz invariance and CPT violations arising from the spontaneous breaking of the underlying space-time symmetry are interesting theoretical feature that can be parametrized within the so-called Standard Model Extension (SME)~\cite{SME19976760,SME1998116002,SME2004105009}. 
Lorentz violating effects in the neutrino sector can appear both in the two-neutrino and in the neutrino-less decay mode~\cite{JORGE2014036002}. Indeed, a distortion of the two-electron summed energy is expected for two neutrino channel due to an extra term in the phase space factor, while neutrinoless could be directly induced by a Lore term.
The signature is very similar to the one expected for Majoron searches with a deformation of the upper part of the two neutrino decay spectrum.

The Pauli exclusion principle (PEP) is one of the basic principles of physics upon which modern atomic and nuclear physics are built. 
Despite its well known success, the exact validity of PEP is still an open question and experimental verification is therefore extremely important~\cite{GREENBERG198983}. 
Indeed, a number of experimental investigations have been carried out both in the nuclear and atomic sector. In all the cases, the signature is a  transition between already occupied (atomic or nuclear) levels which is clearly prohibited by PEP. Most of the low activity experiments exploit large masses and/or low background rates to search for the emission of specific electromagnetic or nuclear radiation from atoms or nuclei~\cite{BXINO2010034317,VIP2018319,BELLI1999236}. Dedicated searches, on the contrary, aim at improving the sensitivity by filling already complete atomic levels with fresh electrons and measuring the corresponding X-ray transitions. Unfortunately a model linking the two experimental observations is still missing and a comparison of the sensitivities is therefore impossible. 
Exploiting an excellent energy resolution and the very low background index it will be possible to look for the emission of X and $\gamma$ rays or of nucleons from the detector atoms and/or nuclei. 

Baryon number (B) conservation is an empirical symmetry of the Standard Model (SM).
Its violation is predicted by a number of SM extensions. Furthermore it is expected that quantum gravity theories violate B and that theories with extra dimensions permit nucleon decay via interactions with dark matter~\cite{BABU20135285}.
In particular, some SM extensions which allow for small neutrino masses, anticipate $\Delta$B=3 transitions in which three baryons can simultaneously disappear from the nucleus, frequently leaving an unstable isotope~\cite{BABU200332}. The coincidence between the tri-nucleon decay and the radioactive decay of the daughter nuclei is then a robust signature which can help to get rid of the backgrounds. The dominant $\Delta$B=3 decay modes are $ppp \to e^+ \pi^+ \pi^+$, $ppn \to e^+ \pi^+$, $pnn \to e^+ \pi^0$, $nnn \to \bar{\nu} \pi^0$. The decay-mode specific signatures (charged fragments) include an initial saturated event followed by one or more radioactive decays while the invisible decay-mode signatures are composed of two successive decays and hence have two energy constraints and one time constraint.

\section{Infrastructure and underground laboratories}
To address fundamental questions in the field of astroparticle physics and nuclear astrophysics at the forefront of research, the access to underground facilities is mandatory in order to mitigate radiogenic and cosmogenic backgrounds and to provide the necessary low-background conditions for DBD0$\nu$-decay searches.

Next-generation DBD0$\nu$-decay experiments, as discussed in section \ref{Status and Prospects}, will apply similar technologies as their precursors however, due to the increase in target mass to ton-scale will under circumstances require/occupy a larger volume. Furthermore, since there is so far no observation of DBD0$\nu$-decay, next generation experiments are part of a world-wide exploratory program necessarily consisting of diverse but complementary techniques using multiple target materials. 
Thus within the next years there will be demand on large/extensive underground space to accommodate all next-generation DBD0$\nu$-decay experiments. This may be realised relying on already existing facilities but also might require the expansion/upgrade of underground laboratories, in particular if there is competition on available space with e.g. future direct dark matter searches and experiments on neutrino physics. 

\subsection{Requirements from Experiments}
Next-generation DBD0$\nu$-experiments will require a careful background evaluation in order to minimize events that could interfere with signal detection, setting them apart from future underground neutrino observatories also regarding the possible choice of the underground site.
Apart from the natural rock overburden at underground sites DBD0$\nu$-experiments are typically surrounding by massive passive shielding (copper, lead) or an active veto (instrumented water tank) to reduce  the background budget. The deeper the underground facility the less shielding is necessary resulting in a smaller footprint of the experimental setup. In this context the depth of the European Laboratori Nazionali del Gran Sasso (LNGS) of INFN with its 3600 m.w.e. (meters of water equivalent) seems to be a good choice satisfying the required depth and, thank to the available underground space and infrastructure, could host all currently proposed next generation DBD0$\nu$-decay European experiments.

Besides the need for cosmic silence, also environmental radioactivity and intrinsic contamination of materials used for detectors, shielding and infrastructure play a key-role for DBD0$\nu$ searches. In this context the possibility for radiopurity assay at underground sites, e.g. material screening relying on existing High Purity Germanium (HPGe) detector screening facilities, as well as  ICP-MS and other high sensitive analytical techniques carried out by experienced and skilled onsite scientists is of great importance to pursue zero-background future DBD0$\nu$-decay searches. For an effective and timely implementation of the experimental programs, pilot experiments will be also needed to design and test complex and costly experimental apparatus. Last but not least, also capacities for storage of detector materials to prevent cosmogenic activation in a radon-free environment will be necessary. To cover and provide all the required supports for the next generation DBD0$\nu$-decay experiments, a strict collaboration between all the European underground laboratories is necessary, their involvement in a close connected European network could be very helpful in supporting the design and construction of next generation experiments.

\subsection{Underground facilities}

Large and deep underground facilities are located in Europe, North America and Asia. In Europe, several underground labs are already hosting (in construction or in data taking) experiments with $\mathcal{O}(100 kg)$ of DBD0$\nu$ isotope. Underground laboratories are currently cooperating in R\&D programs in DBD0$\nu$ and low radioactivity techniques, getting into globally distributed installations. There is e.g. already existing a cooperation between LNGS and LSC on low-background techniques, with the intention to foster and further strengthen the collaboration on complementary material screening services and low radioactivity expertise to support next-generation  DBD0$\nu$ experiments. Here we lay out a possible selection of sites suitable to accommodate next generation DBD0$\nu$-decay searches in the next five to ten years.

\subsubsection{Laboratori Nazionali del Gran Sasso }
The Italian Laboratori Nazionali del Gran Sasso (LNGS) operated by INFN and located about 150 km from Rome is at present the largest running underground site in the world, consisting of three experimental halls (each about 100-meters long, 20-meters large and 18-meters high, 3600 m.w.e.) with a total volume of about 180.000 m{$^3$}. Access to experimental halls is horizontal and via the highway tunnel. LNGS presently already hosts the two DBD0$\nu$-experiments GERDA and CUORE. The laboratory will have capacities to accommodate all the next-generation European DBD0$\nu$ searches \LEG~, CUPID and NEXT-HD/BOLD. The opportunity to install all next-generation experiments at the same site has a couple of attractive and distinct advantages: available infrastructure as e.g. for radiopurity assay as well as expertise on low-background measures may be shared. Same applies for onsite support from experienced and skilled personnel regarding both, infrastructural and scientific support may add an additional value.

\subsubsection{Laboratorio Subterraneo de Canfranc}
The Spanish Laboratorio Subterraneo de Canfrac (LSC) operated by an Spanish Consortium is located in the Pyrenees, in the French-Spanish Somport tunnel, consisting of 1600 m{$^2$} with a total volume of about 10.000 m{$^3$} and a rock overburden equivalent to 2450 m.w.e. It has two experimental halls (40x15x12 m$^3$ and 15x10x7 m$^3$) in which the experiments are distributed as well as offices, a clean room, a mechanical workshop and gas storage room. Access to experimental halls is horizontal and via the highway and the train tunnels. LSC presently already hosts the DBD0$\nu$-experiments NEXT-100, and also CROSS, a demonstrator that studies innovative solutions of the bolometric DBD0$\nu$ technology.

\subsubsection{Laboratoire Subterrain de Modane}
The French Laboratoire Subterrain de Modane (LSM) located in the Fresjus road tunnel near Modane and operated by the French National Center for Scientific Research and the Atomic Energy and Alternative Energies Commission is, with its rock overburden of about 1700 m (4800 m.w.e.) the deepest underground laboratory in Europe. The available volume for the experimental setup installations is about 5000 m{$^3$}. In LSM is installed the SuperNEMO demonstrator detector, which will take data on ${^{82}}$Se double beta decay search, and the CUPID-Mo experiment, currently collecting data with 20 Li$_2$MoO$_4$ scintillating bolometers to investigate the isotope $^{100}$Mo.

\subsubsection{Boulby Underground Laboratory}
Boulby Underground Laboratory (Boulby) is the UK's deep underground science facility, located in working Polyhalite and Salt mine on the North East coast of England. Boulby is funded and operated by the UK’s Science and Technology Research Council (STFC). The Boulby Facility is located 1.1 km below ground, with a {\em flat} overburden of 2820 m w.e. The underground laboratory has a volume of 4,000 m$^{3}$ and operated as class 100,000 clean room with a radiopurity screening area in a class 1,000 room. Confirmed future projects at Boulby include the AIT-WATCHMAN project $-$ a US/UK funded 6~kT Gd-loaded water Cherenkov anti-neutrino detector for nuclear reactor monitoring and technology R\&D. A new cavern of 15,000 m$^{3}$ will be built for WATCHMAN in 2020 with detector operation expected to begin 2024.

\subsubsection{SNOLAB}
The Canadian underground laboratory SNOLAB, located two kilometres below surface (6010 m.w.e.) in the Vale Inco Creighton Mine (near Sudbury), is an expansion of the facility constructed for the SNO solar neutrino experiment. SNOLAB was designed to operate as one large clean room, in keeping with the successful approach used by SNO. The SNOLAB expansion added an additional 6,300 m$^2$ of excavations of which 3,700 m$^2$ is clean room space connected to the existing facility. The clean/dirty boundary was moved for the expanded laboratory and some existing excavations were converted to additional clean space. The SNOLAB underground laboratory has 5,000 m$^2$ of clean space. Of this, 3,100 m$^2$ is experimental laboratory space. There is an additional 2,600 m$^2$ of excavation outside the clean room used by SNOLAB for the service infrastructure and material transportation and storage. SNOLAB has few experimental halls designed for specific experimental requests, in particular the Cryopit hall is equipped to allocate experiments that need large volume of cryogenic liquids.

\subsubsection{China Jingping Underground Laboratory}
CJPL-I was built under Jinping Mountain with 2400 m of rock overburden (6720 m.w.e.). The facility is the deepest operating underground laboratory in the world. It has drive-in access via a two-lane road tunnel with enough headroom for construction trucks. The main hall of CJPL-I, where the experiments were installed, has dimensions of 6.5 m in width, 6.5 m in height  and 40 m in length, and accordingly the floor area is 260 m$^2$. CJPL-II is located 500 m to the west of CJPL-I, along the same road tunnel. It has four caverns, each with dimensions of 14 m in width, 14 m in height and 130 m in length, and is interconnected with access and safety tunnels. The laboratory floor area will be approximately 20000 m$^2$. Two pits will provide additional headroom for specific applications: the first will have a diameter of 18 m and height of 18 m. The second pit will have 27 m in length, 16 m in width, and 14 m in depth. The completion of the underground labs is currently ongoing and all important infrastructures needed 
will be provided in the near future.

\subsubsection{Baksan Underground Laboratory}
The Baksan Neutrino Observatory laboratory is operated by the Institute for Nuclear Research of the Russian Academy of Sciences. It is located under the  Andyrchi mountain in the Northern Caucasus region of Russia. The laboratory has two experimental halls at a shallow and high depths. The shallow hall
(24$\times$24$\times$16)~m$^{3}$  is at 850 m.w.e. and hosted the Baksan Underground Scintillation Telescope. The deep underground hall
(60$\times$10$\times$12)~m$^{3}$  is at 4700 m.w.e. and has hosted the SAGE and BEST radiochemical neutrino detection experiments as well as an HPGe low background facility and a number of R\&D projects on dark matter and DBD0$\nu$-decay. 

\section{Concluding recommendations}


Given the importance of neutrinoless double beta decay searches, the leading role Europe is playing and the prospects for the future, we provide below the key recommendations on the program.

\vspace{0.2truecm}

{\em Recommendation 1. The search for neutrinoless double beta decay
is a top priority in particle and astroparticle physics, as this process provides the most sensitive test of lepton number violation. }

\vspace{0.1truecm}

{\em Recommendation 2. A sustained and enhanced support of the European experimental programme is required to maintain the leadership in the field,  exploiting the broad range of expertise and infrastructure and fostering existing and future international collaborations.}

\vspace{0.1truecm}

{\em Recommendation 3. A multi-isotope program exploiting different technologies at the highest level of sensitivity should be supported in Europe in order to mitigate the risks and to extend the physics reach of a possible discovery.}

\vspace{0.1truecm}

 {\em Recommendation 4. A program of R\&D should be devised on the path towards the meV scale for the effective Majorana mass parameter.}

\vspace{0.1truecm}

{\em Recommendation 5. The European underground laboratories should provide the required space and infrastructure for next generation double beta decay experiments. A strong level of coordination is required among European laboratories for radiopurity material assays and low background instrumentation development in order to ensure that the challenging sensitivities of the next generation experiments can be achieved on competitive timescales.}

\vspace{0.1truecm}

{\em Recommendation 6. The theoretical assessment of the particle physics implications of a positive observation and of the broader physics reach of these experiments should be continued. A dedicated theoretical and experimental effort, in collaboration with the nuclear physics community, is needed to achieve a more accurate determination of the Nuclear Matrix Elements (NME).}

\section*{Acknowledgements}

The committee would like to thank the experimental collaborations, including CUPID, LEGEND, NEXT, DARWIN, nEXO, for providing useful information. We would also like to very much thank J. Menendez, for the explanations of the issues around NME and for providing a first draft of the corresponding section, F. {\v S}imkovic and P. Vogel, for further clarifications and text on these issues, M. Agostini for producing the discovery potential plots  and table and for writing the accompanying text, F. Deppisch for explaining key theoretical issues, W. Rodejohann for highlighting the link between neutrinoless double beta and proton decay. The committee would also like to thank APPEC SAC and GA Chair for valuable feedback and the community for crucial input.

\bibliography{bbreferences}
\end{document}